\documentclass[aps,prb,superscriptaddress,twocolumn,showpacs,amsmath,floatfix]{revtex4-2}

\usepackage{float}
\usepackage{amsmath}
\usepackage{graphicx}
\usepackage{color}
\usepackage{textcomp}
\usepackage{ulem}
\usepackage[colorlinks,linkcolor=blue]{hyperref}



\usepackage{babel}

\begin{document}
\title{Projected entangled pair states study of anisotropic-exchange magnets on triangular lattice}
\author{Meng Zhang}
\affiliation{CAS Key Laboratory of Quantum Information, University of Science and Technology of China, Hefei 230026, Anhui, China}
\affiliation{Synergetic Innovation Center of Quantum Information and Quantum Physics, University of Science and Technology of China, Hefei, 230026, China}
\author{Chao Wang}
\affiliation{CAS Key Laboratory of Quantum Information, University of Science and Technology of China, Hefei 230026, Anhui, China}
\affiliation{Synergetic Innovation Center of Quantum Information and Quantum Physics, University of Science and Technology of China, Hefei, 230026, China}
\author{Shaojun Dong}
\affiliation{CAS Key Laboratory of Quantum Information, University of Science and Technology of China, Hefei 230026, Anhui, China}
\affiliation{Synergetic Innovation Center of Quantum Information and Quantum Physics, University of Science and Technology of China, Hefei, 230026, China}
\author{Hao Zhang}
\affiliation{CAS Key Laboratory of Quantum Information, University of Science and Technology of China, Hefei 230026, Anhui, China}
\affiliation{Synergetic Innovation Center of Quantum Information and Quantum Physics, University of Science and Technology of China, Hefei, 230026, China}
\author{Yongjian Han}%
\email{smhan@ustc.edu.cn}
\affiliation{CAS Key Laboratory of Quantum Information, University of Science and Technology of China, Hefei 230026, Anhui, China}
\affiliation{Synergetic Innovation Center of Quantum Information and Quantum Physics, University of Science and Technology of China, Hefei, 230026, China}
\author{Lixin He}%
\email{helx@ustc.edu.cn}
\affiliation{CAS Key Laboratory of Quantum Information, University of Science and Technology of China, Hefei 230026, Anhui, China}
\affiliation{Synergetic Innovation Center of Quantum Information and Quantum Physics, University of Science and Technology of China, Hefei, 230026, China}

\begin{abstract}
The anisotropic-exchange spin-1/2 model on a triangular lattice has been used
to describe the rare-earth chalcogenides, which may have exotic ground states.
We investigate the quantum phase diagram of the model by using the projected entangled pair state
(PEPS) method, and compare it to the classical phase diagram.
Besides two stripe-ordered phase, and the 120$^\circ$ state, there is also a multi-\textbf{Q} phase.
We identify the multi-\textbf{Q} phase as a $Z_{2}$ vortex state.
No quantum spin liquid state is found in the phase diagram, contrary to the previous DMRG calculations.
\end{abstract}
\maketitle

\section{Introduction}
\begin{ruledtabular}
Over the past few decades, frustrated magnets,
which can host nontrivial order and exotic quantum phases,
have attracted great attentions.
The nearest-neighbor (NN) triangular antiferromagnet is a typical example of
geometric frustration magnets, where the lattice geometry precludes simultaneously
minimization of the energies of all bonds. Anderson proposed that the ground
state of the model is a quantum spin liquid (QSL) state\cite{anderson_resonating_1973,anderson_resonating_1987}.
However, numerical calculations show that the ground state is instead a $120^{\circ}$ order state
\cite{PhysRevLett.60.2531,PhysRevB.50.3108}.
\end{ruledtabular}

The bond-dependent anisotropic exchange can be an alternative approach to enhance
the frustration effects, which may lead to exotic quantum magnetic states.
A famed example is the Kitaev spin model on the honeycomb lattice \cite{kitaev_anyons_2006}, where each bond with different spatial orientation carries a different Ising-like interaction. An exact solution of this model
shows that the ground state is a QSL state with fractional spin excitations.
Very recently, a large family of rare-earth triangular lattice materials, including YbMgGaO$_{4}$ \cite{li_gapless_2015,Li_2015,li_rare-earth_2015,li_muon_2016,li_crystalline_2017,shen_evidence_2016,paddison_continuous_2016,li_rearrangement_2019} and rare-earth chalcogenide family NaYbCh$_{2}$ (Ch = O, S, Se) \cite{liu_rare-earth_2018,bordelon_field-tunable_2019},
have been synthesized and explored experimentally. In these materials, Yb$^{3+}$ ions form perfect triangular layer with effective spin-1/2 local moments. Strong spin-orbit coupling (SOC) introduces the anisotropic spin interactions,
which provides ideal platforms to explore the interplay of
 bond-dependent exchange frustration and geometric frustration.
Experimentally \cite{li_gapless_2015,Li_2015,li_rare-earth_2015,li_muon_2016,
	li_crystalline_2017,shen_evidence_2016,paddison_continuous_2016,
	li_rearrangement_2019,liu_rare-earth_2018,bordelon_field-tunable_2019}, it has been demonstrated that these rare-earth materials show no long-range magnetic order down to low temperature, which have been proposed as promising candidates as quantum spin liquids.

To understand the magnetic properties of the triangular materials,
a generic spin-1/2  spin model with anisotropic exchange interactions
on triangular lattice based on symmetry consideration
has been proposed in Ref. \cite{li_gapless_2015,li_anisotropic_2016}.
This model reduces to the Kitaev-Heisenberg (KH) model for certain exchange parameters.
Despite the model has been studied intensively by various numerical and theoretical methods,
including the classical spin method \cite{maksimov_anisotropic-exchange_2019,becker_spin-orbit_2015,rousochatzakis_kitaev_2016,liu_semiclassical_2016}, mean-field theory
\cite{PhysRevB.95.024421,Li_2015}, exact diagonalization (ED) \cite{wu_exact_2020} and
density matrix renormalization group (DMRG) method \cite{maksimov_anisotropic-exchange_2019,zhu_topography_2018,luo_ground-state_2017},
the phase diagram of the model is still illusive.
Especially, DMRG calculations \cite{maksimov_anisotropic-exchange_2019,zhu_topography_2018},
as well as ED \cite{wu_exact_2020} calculations on small clusters,
show that there is a QSL region between the stripe phases and the 120$^\circ$ phase. However,
the QSL state calculated by ED has a different parent state than that of the DMRG result.
Meanwhile other DMRG \cite{shinjo_density-matrix_2016} studies on the KH model show that this
region in the phase diagram actually has a $Z_{2}$ vortex ground state, instead of QSL.
It is therefore very urgent and important to know whether there is indeed a QSL region in the phase diagram.

To clarify the phase diagram of the anisotropic exchange spin-1/2 model on triangular lattice,
we perform high accuracy projected entangled pair states (PEPS) calculations,
and compare the results to those of the classical simulations.
We explore the phase diagram of the model, and find the stripe
phases, $120^{\circ}$ phase, and multi-\textbf{Q} phase in both quantum
and classical models.
Most importantly, we identify that the multi-\textbf{Q} phase, between
the stipe orders and the $120^{\circ}$ phase,
is a $Z_2$ vortex state, which would NOT melt to QSL by quantum fluctuation.

The rest of the paper is organized as follows. In Sec.~\ref{sec:methods}, we introduce
the anisotropic exchange model on triangular lattice, and the methods used in this work.
We present the phase diagram of the classical simulations in Sec.~\ref{sec:classical}
and the phase diagram of the quantum model
in Sec.~\ref{sec:quantum}. We discuss more on the $Z_2$ vortex state in Sec.~\ref{sec:discussion} and
summarize the work in Sec.\ref{sec:summary}

\section{Model and methods}
\label{sec:methods}

\begin{figure}[tb]
\includegraphics[scale=0.27]{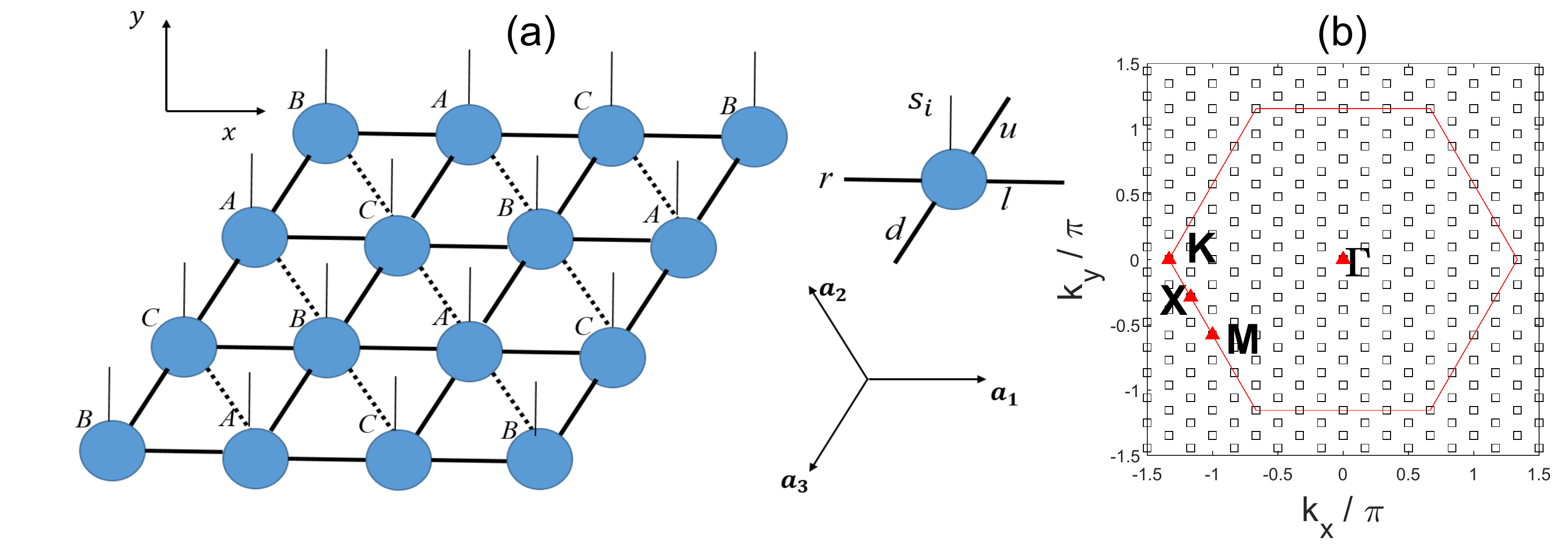}
\caption{\label{fig1}
	(a) A schematic diagram of PEPS on a triangular lattice.
The thick solid lines represent the virtual bonds of the PEPS, whereas the thin vertical lines
are the physical indices of the PEPS. ``$A$'', ``$B$'' and ``$C$'' in the figure denote
the three sublattices of triangular lattice. $\varphi_{\alpha}=\left\{ 0,2\pi/3,-2\pi/3\right\} $ for
	the bond along the primitive vectors $\mathbf{a}_{1}$, $\mathbf{a}_{2}$ and $\mathbf{a}_{3}$, respectively.
	(b) The Brillouin zone of the triangular lattice, where the squares denote the $\mathbf{k}$ points of a $12\times12$ lattice.
$\Gamma$, $K$, $M$ are the high symmetric points in the Brillouin zone, and
$X$ point is located on the line between the $K$ and $M$ points.}

\end{figure}

The generic nearest-neighbor spin-1/2 Hamiltonian, which is invariant
under symmetry group $R\overline{3}m$,
can be written as~\cite{li_gapless_2015,li_anisotropic_2016},
\begin{multline}
\mathcal{H}=\sum_{\left\langle ij\right\rangle }\left\{ J\left(S_{i}^{x}S_{j}^{x}+S_{i}^{y}S_{j}^{y}+\Delta S_{i}^{z}S_{j}^{z}\right)\right.\\
+2J_{\pm\pm}\left[\left(S_{i}^{x}S_{j}^{x}-S_{i}^{y}S_{j}^{y}\right)\cos\varphi_{\alpha}-\left(S_{i}^{x}S_{j}^{y}+S_{i}^{y}S_{j}^{x}\right)\sin\varphi_{\alpha}\right]\\
\left.+J_{z\pm}\left[\left(S_{i}^{y}S_{j}^{z}-S_{i}^{z}S_{j}^{y}\right)\cos\varphi_{\alpha}-\left(S_{i}^{x}S_{j}^{z}+S_{i}^{z}S_{j}^{x}\right)\sin\varphi_{\alpha}\right]\right\} \label{eq:(1)}
\end{multline}
Here $\mathbf{S}_{i}$ refers to the spin-1/2 operators on site $i$,
and $\left\langle ij\right\rangle $ denotes a pair of nearest neighbor spins.
$\varphi_{\alpha}=\left\{ 0,2\pi/3,-2\pi/3\right\} $
are the angles between the lattice vectors $\mathbf{a}_{\alpha}$ and the $x$-axis [see Fig.~\ref{fig1}(a)]
in crystallographic axes.
The first term of Eq.(\ref{eq:(1)}) is the standard XXZ model and is invariant under the global spin rotation around the $z$-axis. The $J_{\pm\pm}$ and $J_{z\pm}$ terms define the bond dependent anisotropic interactions caused by the strong SOC,
which break the rotational symmetries of Hamiltonian, but retain the time reversal symmetry.
In this work, we set $J$=1, and $J_{z\pm}$ take positive values, because the phase diagram for $-J_{z\pm}$ can be obtained by a global $\pi$ rotation around the $z$ axis \cite{li_anisotropic_2016}.

It is also natural to rewrite model (\ref{eq:(1)}) in the cubic axes.
When $\Delta=1$ and $J_{z\pm}=2\sqrt{2}J_{\pm\pm}$, model
(\ref{eq:(1)}) is reduced to the well-known KH model
in the cubic axes \cite{maksimov_anisotropic-exchange_2019}.
More details of relation between model (\ref{eq:(1)}) and the KH model are given in Appendix \ref{cubic}.
Recent first-principles calculations show that $\Delta$ is rather close to 1 in NaYbS$_2$
($\Delta$=0.980) and NaYbO$_2$ ($\Delta$=0.889  ) \cite{zheng2021firstprinciples}.
Therefore, in this work, we focus on the phase diagram in the case of $\Delta=1$.

We investigate the anisotropic triangular model (\ref{eq:(1)}),
by using both classical method and quantum many-particle method.
In the classical simulations, the spins are treated as unimodular classical vectors.
We obtain the ground state of the classical model on a 30 $\times$ 30 lattice by optimizing the total energies via
a replica exchange molecular spin dynamics method \cite{PhysRevLett.57.2607,liu_replica_2015,PhysRevB.96.115119},
which can effectively avoid the system being trapped in the local minima.

To investigate the phase diagram of the quantum model, we employ the PEPS method.
The ground state wave functions are presented by PEPS on the $N=L_{x}\times L_{y}$
triangular lattices with open boundary conditions, as schematically shown in Fig.\ref{fig1}(a),
\begin{equation}
\left|\Psi\right\rangle =\sum_{s_{1}\cdots s_{N}=1}^{d}Tr(A_{1}^{s_{1}}A_{2}^{s_{2}}\cdots A_{N}^{s_{N}})\left|s_{1}s_{2}\cdots s_{N}\right\rangle
\end{equation}
where tensor $A_{i}^{s_{i}}=A_{i}(r,l,u,d,s_{i})$ is a five-index
tensor located on site $i$. $s_{i}$ is the physical index and $r$, $l$ ,$u$, $d$
are the virtual bonds of the PEPS, with a bond dimension $D$.
In this study, all results are obtained by the PEPS with $D$=8, unless otherwise noted.
To obtain highly accurate ground state,
the PEPS wave functions are first optimized by the imaginary time evolution with a
simple update method \cite{PhysRevLett.101.090603},\
followed by a stochastic gradient optimization method \cite{PhysRevB.95.195154}.

To distinguish different ordered states, we calculate the spin structure factor (SSF) for each spin component,
\begin{equation}
	S^{\nu}\left(\mathbf{q}\right)=\frac{1}{N}\sum_{ij}e^{i\mathbf{q.\left(R_{i}-R_{j}\right)}}\left\langle S_{i}^{\nu}S_{j}^{\nu}\right\rangle \, ,
\end{equation}
where $\nu=x, y, z$ are the spin components and $N$ is the total number of sites. The total SSF is the sum of the
three components.
In Fig.~\ref{fig1}(b), we show the Brillouin zone of the $12\times12$ triangular lattice,
where $X$ point is located on the line between the $K$ and $M$ points.

\section{Results and Discussion}

\subsection{Phase diagram of the classical model}\label{sec:classical}

\begin{figure}[tb]
	\includegraphics[scale=0.58]{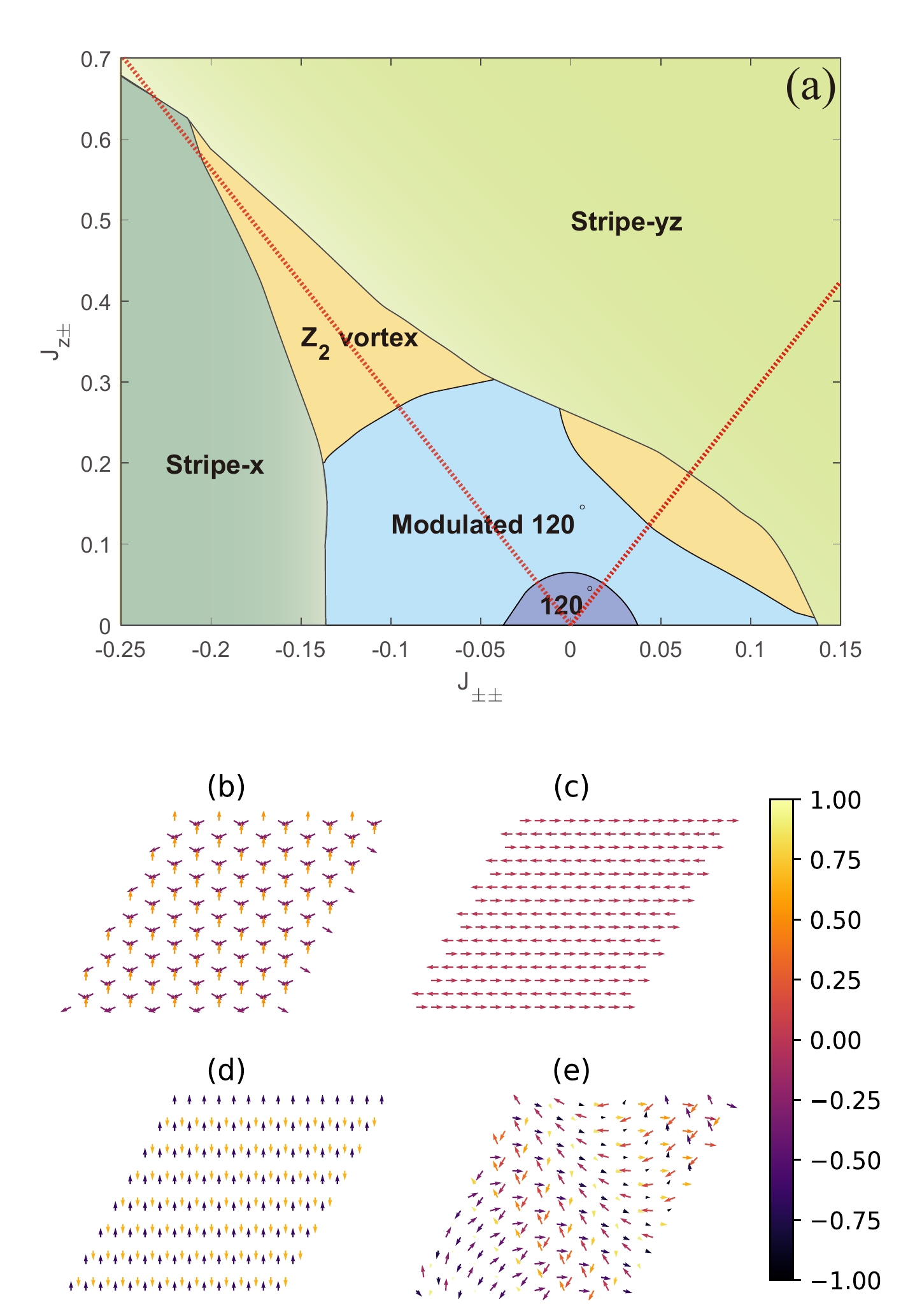}	
	\caption{\label{spin_config}
		(a) Classical phase diagram for model \eqref{eq:(1)}.
		On the red dotted line ($J_{z\pm}=\pm2\sqrt{2}J_{\pm\pm}$),
the anisotropic spin model \eqref{eq:(1)}, reduces to the KH model.
(b)-(e) The spin configurations of (b)the $120^{\circ}$ phase,
		(c)the stripe-$x$ phase, (d)the stripe-$yz$ phase,
and (e)the  modulated $120^{\circ}$ phase. The arrows correspond to
the projections of the classical spins onto the $x$-$y$ plane, whereas the color of spins
indicates the magnitude of $S_{z}$.}
\end{figure}

The phase diagram of the classical simulations in the $J_{z\pm}$-$J_{\pm\pm}$ plane is shown in Fig.\ref{spin_config} (a)
for $\Delta$=1.
There are five different phases.
On the left side of the phase diagram, the ground state is of a stripe-$x$ order, where the spins lie
in the $x$-$y$ plane. In the top-right corner of the phase diagram, there is another stripe-ordered phase,
the stripe-$yz$ phase, where the spins are partially out of the $x$-$y$ plane.
In the vicinity of  $J_{z\pm}$, $J_{\pm\pm}$ $\sim$0, there is the 120$^{\circ}$ phase.
The spin textures of the 120$^{\circ}$, stripe-$x$ and stripe-$yz$ phases are shown in Fig.\ref{spin_config} (b)-(d),
respectively. All these states are single-\textbf{Q} commensurate states.

In the intermediate region of the phase diagram, multi-\textbf{Q} phases become
the ground states, in which the magnetic moments are ordered at multiple \textbf{Q} vectors
in the Brillouin zone.
One of the multi-\textbf{Q} phases, which surrounds the
$120^{\circ}$ phase, is the modulated $120^{\circ}$ phase.
The non-coplanar spin configuration of modulated $120^{\circ}$ phase
remains close to the $120^{\circ}$ configuration locally, but modulates at larger
scale, as shown in Fig.\ref{spin_config}(e). The typical total SSF of the (classical)
modulated $120^{\circ}$ phase is given in Fig.\ref{fig:sf_m120}(a) of the Appendix~\ref{sec:SFs},
which shows that the peaks of the total SSF move slightly away from the
wave vector $K$ point, which is the ordering vector of the 120$^{\circ}$ state.

\begin{figure*}[t]
\includegraphics[scale=0.6]{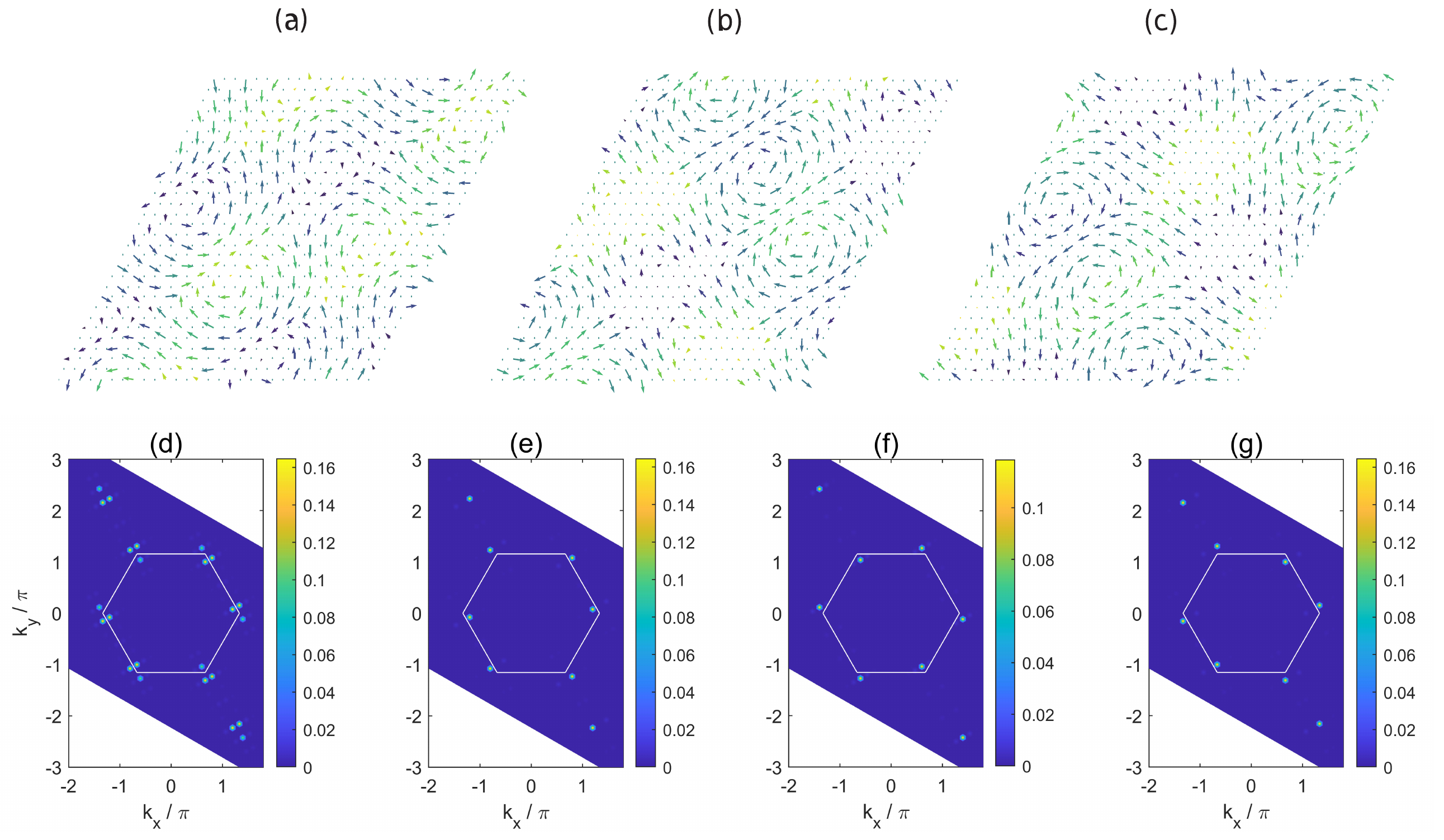}
\caption{\label{Classical result}
	Spin configurations and SSFs of the classical $Z_{2}$ vortex phase.
   (a)-(c) The spin configurations of three sublattices, where the spins have been projected onto the $x$-$y$ plane, and the yellow arrows point upwards out of the plane and  blue arrows point downwards out of the plane.
	(d) The total SSF of the $Z_{2}$ vortex phase, calculated at $J_{\pm\pm}=-0.1,J_{z\pm}=0.325$ in the cubic axes, and (e)-(g) the SSFs of the $x$, $y$ and $z$ components.}
\end{figure*}

In addition to the modulated $120^{\circ}$ phase, there is another multi-\textbf{Q} phase,
which can be identified as a $Z_{2}$ vortex state by its real-space spin textures
and SSFs.
In Fig.\ref{Classical result} (a)-(c), we plot
the spin configurations in the three sublattices
of a $30\times30$ lattice. The spins configurations of a given sublattice
consist of ferromagnetic (FM) domains and vertices. In the region,
where one sublattice forms a FM domain,
the spins of other two sublattices form vertices.
More specifically, when we trace
a closed path around the center of the FM domain,
the spins of other two sublattices complete a $2\pi$ rotation
around the center, i.e., the spin texture forms $Z_{2}$ vortices.
The parent state of $Z_{2}$ vortex state is also
the $120^{\circ}$ state, who has similar local structure to the $120^{\circ}$ state ,
but form vortex on large scale.

When $\Delta=1$ and $J_{z\pm}=\pm 2\sqrt{2}J_{\pm\pm}$
[i.e., the two red lines in Fig.\ref{spin_config} (a)], model(\ref{eq:(1)})
reduces to the KH model in the cubic axes \cite{maksimov_anisotropic-exchange_2019},
which has been extensively studied by various numerical methods \cite{becker_spin-orbit_2015,rousochatzakis_kitaev_2016,li_collective_2019}.
It has been shown that there is a $Z_{2}$ vortex state in both positive and negative
Kitaev interactions region of KH model, which corresponds to the
positive and negative $J_{\pm\pm}$ regions in Fig.\ref{spin_config} (a).
To compare our results with those of the KH model,
we transform the spin configuration of the $Z_2$ vortex phase
from the crystallographic reference frame to the cubic frame \cite{maksimov_anisotropic-exchange_2019}
and recalculate the SSFs of different spin components.
Figure~\ref{Classical result}(d) shows the total SSF,
whereas Fig.\ref{Classical result} (e)-(g) show the SSFs of $x$, $y$ and $z$ spin components, respectively.
The peaks of the SSFs of the three spin components move slightly away from the
$K$ points along different directions.
The SSFs are essentially the same with those of the $Z_{2}$ vortex phase in the KH model \cite{becker_spin-orbit_2015,rousochatzakis_kitaev_2016}.
The results suggest that the $Z_2$ state is stable in a quite large parameter space
beyond the KH model.

The anisotropic triangular spin model has been studied via
the instabilities of magnons~\cite{maksimov_anisotropic-exchange_2019}.
Reference~\cite{maksimov_anisotropic-exchange_2019}
also finds a multi-\textbf{Q} phase in the case of $\Delta$=1,
in which the magnon spectra are soften at three symmetric $\mathbf{k}$ points in the
immediate vicinity of the $\Gamma$ and $K$ points.
The phase has been identified as the $Z_2$ phase.
However, in Ref.~\cite{maksimov_anisotropic-exchange_2019}, the multi-\textbf{Q} phase
is proposed as a single $Z_{2}$ vortex state, in contrast to our results that there is also
a modulated 120$^{\circ}$ state. Furthermore, in Ref. \cite{maksimov_anisotropic-exchange_2019},
the magnons of the $120^{\circ}$ phase at $\Delta=1$ are unstable to any finite value of $J_{z\pm}$,
which is in contrast to our results that $120^{\circ}$ phase is stable when
the anisotropic interactions $J_{z\pm}$ and $J_{\pm\pm}$ are small.

\subsection{Phase diagrams of quantum model}
\label{sec:quantum}

\begin{figure}[tb!]
	\includegraphics[scale=0.62]{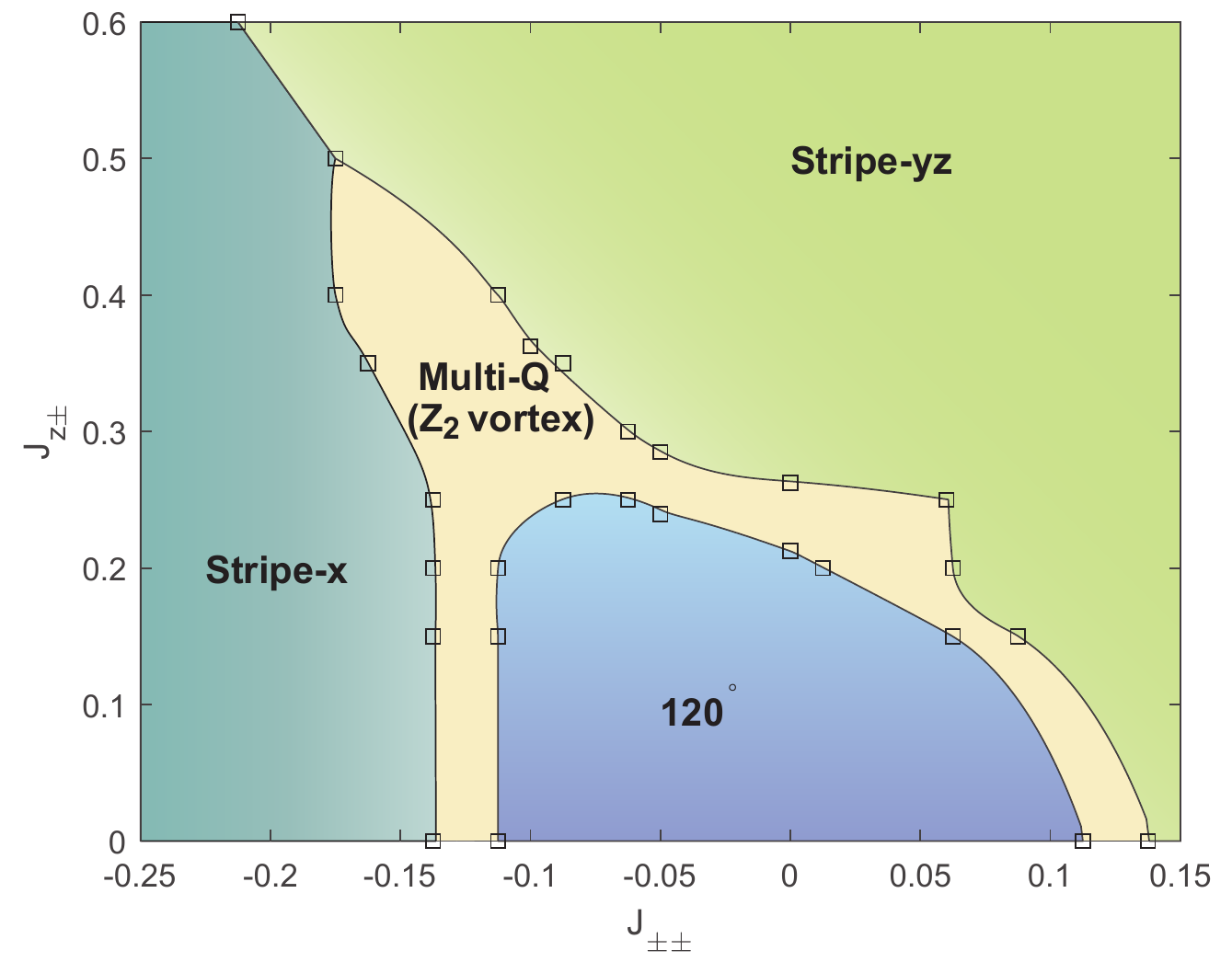}
	\caption{\label{quamtum phase}
	The	quantum phase diagram for model \eqref{eq:(1)}, including the strip-$x$, stripe-$yx$, 120$^\circ$ and
a $Z_2$ vortex phase.}
	
\end{figure}

In this section, we investigate the phase diagram of the anisotropic triangular model
quantum mechanically  to understand the effects of quantum fluctuation.
The phase diagram is shown in Fig.~\ref{quamtum phase} for $\Delta=1$.
Similar to the classical model, the stripe-$x$ phase and the stripe-$yz$ phase
are on the bottom-left and top-right corner respectively.
In the intermediate region, there is a 120$^{\circ}$ phase and a multi-\textbf{Q} phase.

The phase diagram is determined by the order parameters $M(\mathbf{Q})$=$\sqrt{\sum_{\upsilon}S^{\nu}\left(\mathbf{Q}\right)/N}$,
where $S^{\nu}$ ($\nu=x,y,z$) is the $\nu$-th SSF of spin component, and $\mathbf{Q}$ is the spin wave vector.
The typical total SSFs for the $120^{\circ}$, the stripe and the  multi-\textbf{Q}  states on the 12$\times$12 lattice are shown in Fig.~\ref{fig:three oreder}.
The SSF of the $120^{\circ}$  phase have peaks at the  $\mathbf{Q}$=$K$ point, whereas the SSF of the stripe phases
have peaks at the $M$ point. The SSF of a multi-\textbf{Q} phase has peaks around the $X$ point, which lies in line between the $M$ and $K$ points.
The details to determine the phase boundaries are given in Appendix \ref{sec:AppB}.

Since the stripe orders are trivial, and therefore will not be discussed further in the paper.
We will focus on the multi-\textbf{Q} phase and the 120$^{\circ}$ phase in more details.

\begin{figure}[tb!]
\includegraphics[scale=0.5]{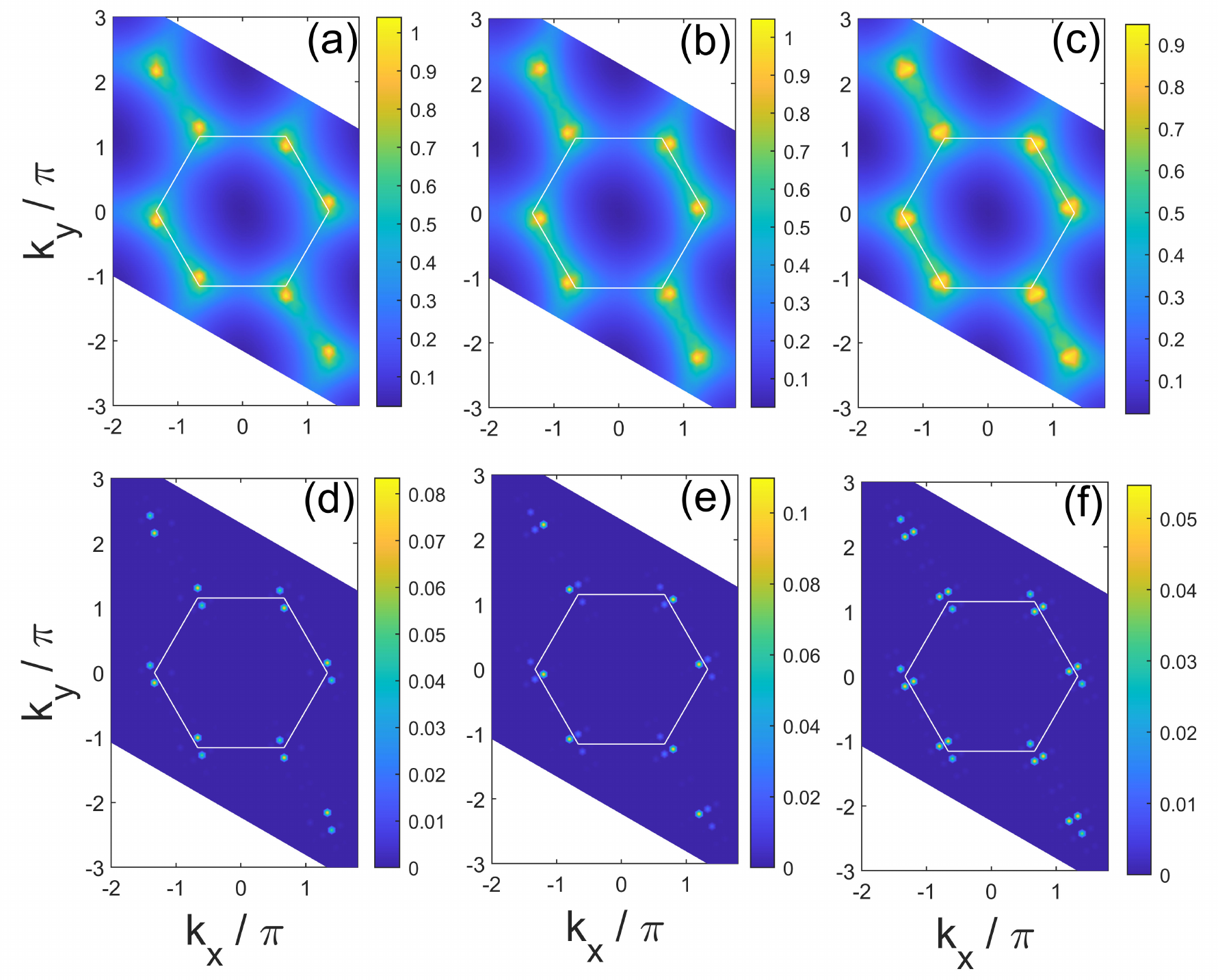}
\caption{\label{Comparing SF}
The SSFs of the quantum and classical $Z_{2}$ vortex phases in the crystallographic axes.
(a)-(c) The SSFs of the $x$, $y$, and $z$ components of the quantum spin model at $J_{\pm\pm}=-0.1,J_{z\pm}=0.325$, respectively.
(d)-(f) The corresponding classical SSFs of the $x$, $y$, and $z$ components. }
\end{figure}

{\it multi-\textbf{Q} ($Z_2$ vortex) phase}: In Sec.~\ref{sec:classical}, we identify that the  multi-\textbf{Q} phase
of the classical model with large anisotropic interaction is a $Z_2$ vortex state.
To further determine the nature of the quantum model, we plot the SSFs of
the quantum model in Fig.\ref{Comparing SF}(a)-(c)
for the $x$, $y$ and $z$ spin components respectively, on a 15$\times$15 lattice.
We compare the SSFs with those of the classical model shown in Fig.\ref{Comparing SF}(d)-(f).
Both results are computed in the crystallographic reference frame.
The SSFs of the $x$, $y$ and $z$ components show that the primary peaks of the quantum model
are essentially same with those of the classical model, suggesting that
this multi-\textbf{Q} phase is also a $Z_{2}$ vortex state.

\begin{figure}[tb!]
\includegraphics[scale=0.4]{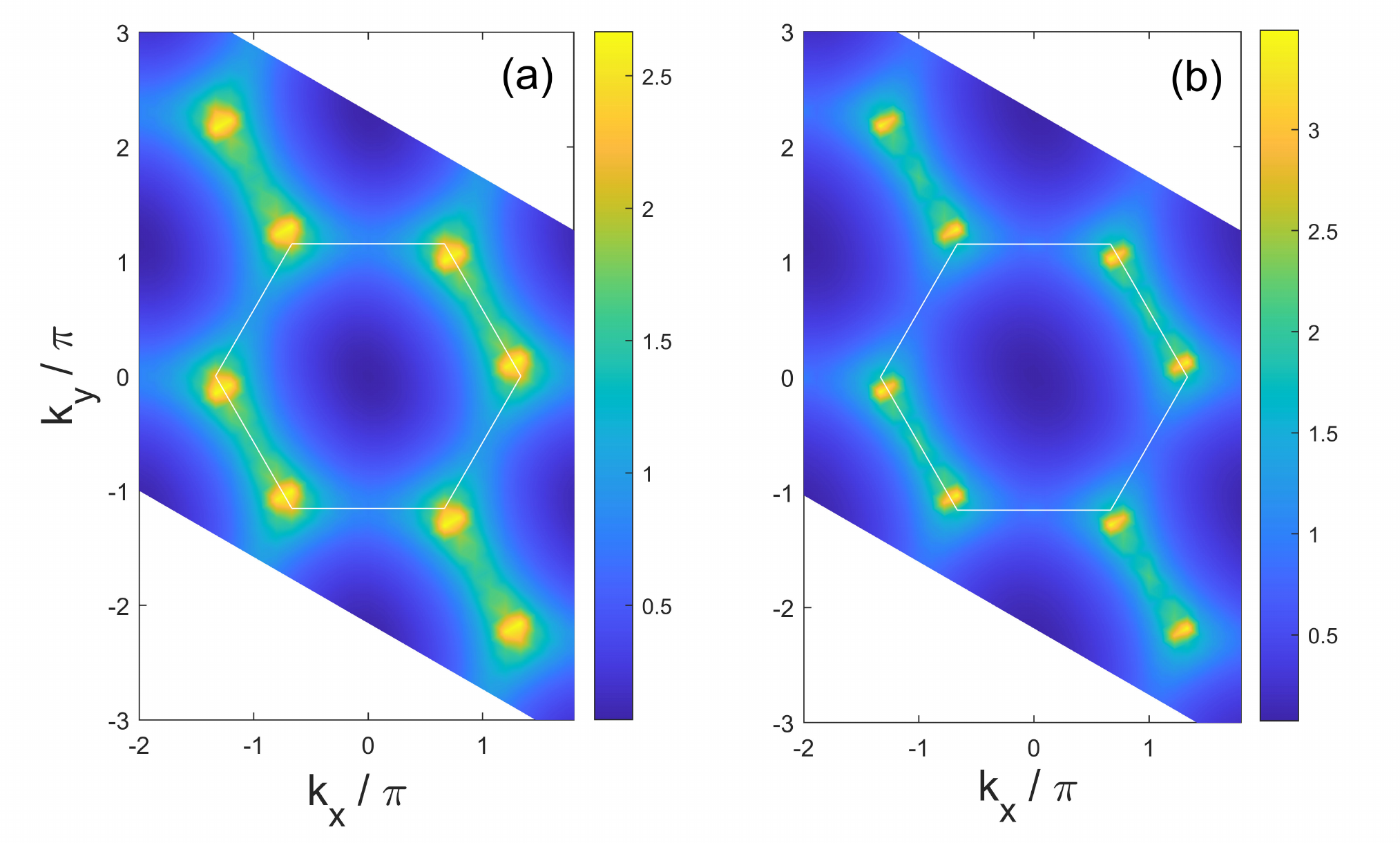}
\caption{\label{Z2 SF 15 18}
	The total SSFs of the quantum spin model at $J_{\pm\pm}=-0.1,J_{z\pm}=0.325$
	on the (a) $15\times15$ lattice and (b) $18\times18$ lattice.}
\end{figure}

We note that previous DMRG calculations find a possible QSL phase in the region of $J_{z\pm}\simeq$[0.27,0.45]
and $J_{\pm\pm}\simeq$[-0.17,0.1] in the isotropic limit $\Delta$=1
\cite{zhu_topography_2018,maksimov_anisotropic-exchange_2019},
in contrast to the $Z_{2}$ vortex state obtained in this work.
We note that the DMRG calculations were performed on rather narrow cylinders.
To further check the stability of $Z_{2}$
vortex phase in this region, we perform more detailed calculations using $J_{\pm\pm}$=-0.1, $J_{z\pm}$=0.325,
which is in the center of the possible QSL phase in previous DMRG studies \cite{zhu_topography_2018},
on larger lattices.
Figure~\ref{Z2 SF 15 18}(b) and (c)
depict the total SSFs for the 15$\times$15 and 18$\times$18 lattices, respectively.
As we see, the SSF show even sharper peaks around the $K$ points for larger lattices,
which suggests that the $Z_2$ phase is stable and does not show any
traces of melting in larger systems. Therefore, we conclude that model(\ref{eq:(1)})
has no QSL ground state or has only very small QSL region in the phase
diagram.


{\it 120$^{\circ}$ phase}: The most noticeable difference between the phase diagrams of the classical model and quantum model, is that the modulated $120^{\circ}$ phase disappears in the quantum model,
and the region of the $120^{\circ}$ phase are much larger than that of the classical model.
The (classical) spin waves results also show that magnons in the $\Delta=1$ limit are unstable to any finite value of $J_{z\pm}$, but the DMRG results show that the $120^{\circ}$ state
is  stable for $J_{z\pm}\leq0.27$ \cite{maksimov_anisotropic-exchange_2019}.
However, we find that the $120^{\circ}$ state in the large anisotropic interactions $J_{z\pm}$ and $J_{\pm\pm}$ region,
becomes unstable for larger systems.
In Fig \ref{fig:sf_m120}(b)(c), we compare the total SSFs of the $12\times12$ and  $15\times15$ lattices
for $J_{\pm\pm}$=-0.025 and $J_{z\pm}$=0.225.
Comparing to the SSF of the $12\times12$ lattice,
the primary peaks of the $15\times15$ lattice move away slightly from the $K$ point.
The results suggest that the region of the 120$^{\circ}$ phase region in thermodynamic limit
is smaller than that shown in Fig.\ref{quamtum phase}, calculated on the $12\times12$ lattice.

\subsection{Discussion}
\label{sec:discussion}

The $Z_{2}$ vortex state is a distortion of its $120^{\circ}$ parent state.
The interplay of bond-dependent anisotropic exchange and geometric frustration is
the key to induce the $Z_{2}$ vortex phase \cite{PhysRevB.101.174443}.
In the KH model, distance between vortices increases with decreasing
anisotropic Kitaev interaction. Therefore, in relatively small systems,
the area of $Z_{2}$ vortex phase is reduced,
which suggests that large system size is necessary to detect
the $Z_{2}$ vortex state.

Our calculations show that the  $Z_{2}$ vortex state is stable in the intermediate
region between $120^{\circ}$ phase and the two stripe phases, and the quantum
fluctuations would not destroy the classical $Z_{2}$ vortex state.
This is contrary to the DMRG results which find a QSL phase in the region \cite{zhu_topography_2018,maksimov_anisotropic-exchange_2019}.
In the anisotropic triangular model, the large coordination number play vital roles to stabilize
the classical magnetic orders, despite bond-dependent exchange frustration and geometric
frustration coexist. \cite{wang_comprehensive_2020}.

The $Z_{2}$ vortex state can be detected by small-angle neutron or x-ray scattering methods and NMR.
There are many materials that can be described by model (1), including rare-earth chalcogenide
family NaYbCh$_{2}$ (Ch = O, S, Se) \cite{liu_rare-earth_2018,bordelon_field-tunable_2019},
YbMgGaO$_{4}$ \cite{li_gapless_2015,Li_2015,li_rare-earth_2015,li_muon_2016,li_crystalline_2017,shen_evidence_2016,paddison_continuous_2016,li_rearrangement_2019}
and iridate Ba$_{3}$IrTi$_{2}$O$_{9}$, etc. These materials can server as a potential platforms to
observe the $Z_{2}$ vortex state.

We note that some experimental studies have found some spin liquid signature
in these materials, but our results suggest that the nearest-neighbor interaction model in triangular lattice
do not give rise to spin liquid phase.
Additional interactions might be
important for describe these material, for instance next nearest-neighbor interaction
\cite{zheng2021firstprinciples} or the inter-layer interactions
\cite{bordelon_field-tunable_2019}.

\section{Summary}
\label{sec:summary}

We investigate the phase diagram of the quantum anisotropic triangular model by using
the PEPS method, and compare it to that of the classical model.
The results show the quantum phase diagram is very similar to that of the classical model.
We identify the multi-\textbf{Q} phase between the stripe phases and the 120$^\circ$ phase
as the $Z_{2}$ vortex state, which is stable even under the of quantum fluctuation.
No quantum spin liquid state is found in the phase diagram, contrary to the previous DMRG calculations.

\acknowledgments

The work is supported by the National Science Foundation of China (Grant Number 11774327, 11874343).
The numerical calculations have been done on the USTC HPC facilities.

\appendix

\section{}
\subsection{Kitaev-Heisenberg model}\label{cubic}

Many rare-earth magnets can support the Kitaev type of bond dependent exchange interactions, including the materials of honeycomb lattice, triangular lattice and pyrochlore lattice built from edge-sharing octahedra \cite{PhysRevB.98.054408}. In the rare-earth compounds, such as YbMgGaO$_{4}$ and NaYbCh$_{2}$ (Ch = O, S, Se),
the layered triangular lattice structures are composed of magnetic ions surrounded by edge sharing octahedra of ligands. The bonds of magnetic ions and ligands form cubic shapes \cite{maksimov_anisotropic-exchange_2019,PhysRevB.98.054408}. Following the choice of Ref. \cite{maksimov_anisotropic-exchange_2019}, the transformation from the cubic to crystallographic reference frame, ${\bf S}_{\rm cryst}\!=\!
\hat{\mathbf{R}}{\bf S}_{\rm cubic}$, is given by

\begin{align}
	\hat{\mathbf{R}}=\left(
	\begin{array}{ccc}
		0 & \frac{1}{\sqrt{2}} & \frac{1}{\sqrt{2}} \\
		-\sqrt{\frac{2}{3}} & \frac{1}{\sqrt{6}} & -\frac{1}{\sqrt{6}} \\
		-\frac{1}{\sqrt{3}} & -\frac{1}{\sqrt{3}} & \frac{1}{\sqrt{3}} \\
	\end{array}
	\right).
	\label{transform}
\end{align}
The Hamiltonian \eqref{eq:(1)} can be rewritten as
the extended Kitaev-Heisenberg model:
\begin{align}
	\mathcal{H}=\sum_{\langle ij \rangle_\gamma}& \Big[
	J_0 \mathbf{S}_i \cdot \mathbf{S}_j +K S^\gamma_i S^\gamma_j
	+\Gamma \left( S^\alpha_i S^\beta_j +S^\beta_i S^\alpha_j\right)\nonumber\\
	+&\Gamma' \left( S^\gamma_i S^\alpha_j+S^\gamma_i S^\beta_j+S^\alpha_i S^\gamma_j
	+S^\beta_i S^\gamma_j\right)\Big],
	\label{JK_gene}
\end{align}
where $\{\alpha,\beta,\gamma\}$ = $\{ y,z,x \}$, $\{ z,x,y \}$,
and $\{ x,y,z \}$, for the ${\rm X}$ bond, ${\rm Y}$ bond, and ${\rm Z}$ bond, respectively,
and $\{ \text{X,Y,Z}\}\equiv \{ \pm\mathbf{a}_{1},\pm\mathbf{a}_{2},\pm\mathbf{a}_{3}\}$. In the $\Delta$=1 limit,
and along the line of $J_{z\pm}\!=\!2\sqrt{2}J_{\pm \pm}$, model \eqref{JK_gene} reduces to the Kitaev-Heisenberg model,
\begin{align}
	\mathcal{H}=\sum_{\langle ij \rangle_\gamma}
	J_0 \mathbf{S}_i \cdot \mathbf{S}_j +K S^\gamma_i S^\gamma_j\, ,
	\label{JK_line}
\end{align}
where $J_0\!=\!J+2J_{\pm\pm}$ and $K\!=\!-6 J_{\pm\pm}$.

\subsection{Phase boundary of the quantum model}
\label{sec:AppB}

\begin{figure}[tb!]
	\includegraphics[scale=0.38]{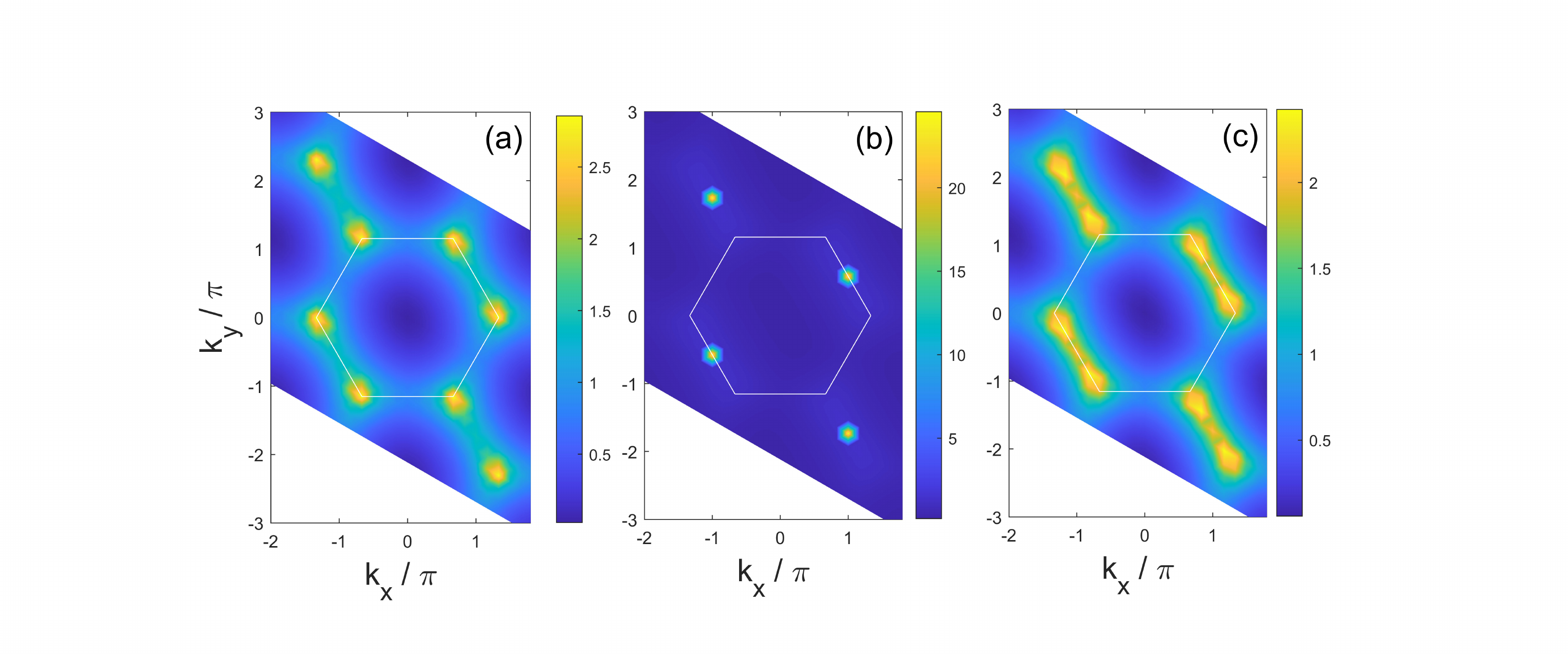}
	\caption{\label{fig:three oreder}
		Typical total SSFs of the quantum model of (a) the $120^{\circ}$ phase at $J_{\pm\pm}=0,J_{z\pm}=0.15$; (b) the stripe phase-$yz$ at $J_{\pm\pm}=0.2,J_{z\pm}=0.2$,
	and	(c) the  multi-\textbf{Q} phase for $J_{\pm\pm}=-0.1,J_{z\pm}=0.25$. The results are calculated on a $12\times12$ lattice.}
\end{figure}

The typical total SSFs of the $120^{\circ}$ phase, stripe phases and multi-\textbf{Q} phase
 are shown in Fig.~\ref{fig:three oreder} for the quantum spin model.
The SSFs are calculated on a 12$\times$12 lattice via PEPS method.
The SSFs are peaked at the $K$ point, $M$ point, and around the $X$ point
for the $120^{\circ}$ phase, stripe phases and multi-\textbf{Q} phases, respectively.
The values of the peaks can be used as the order parameters of the phases.

\begin{figure}[tb!]
	\includegraphics[scale=0.68]{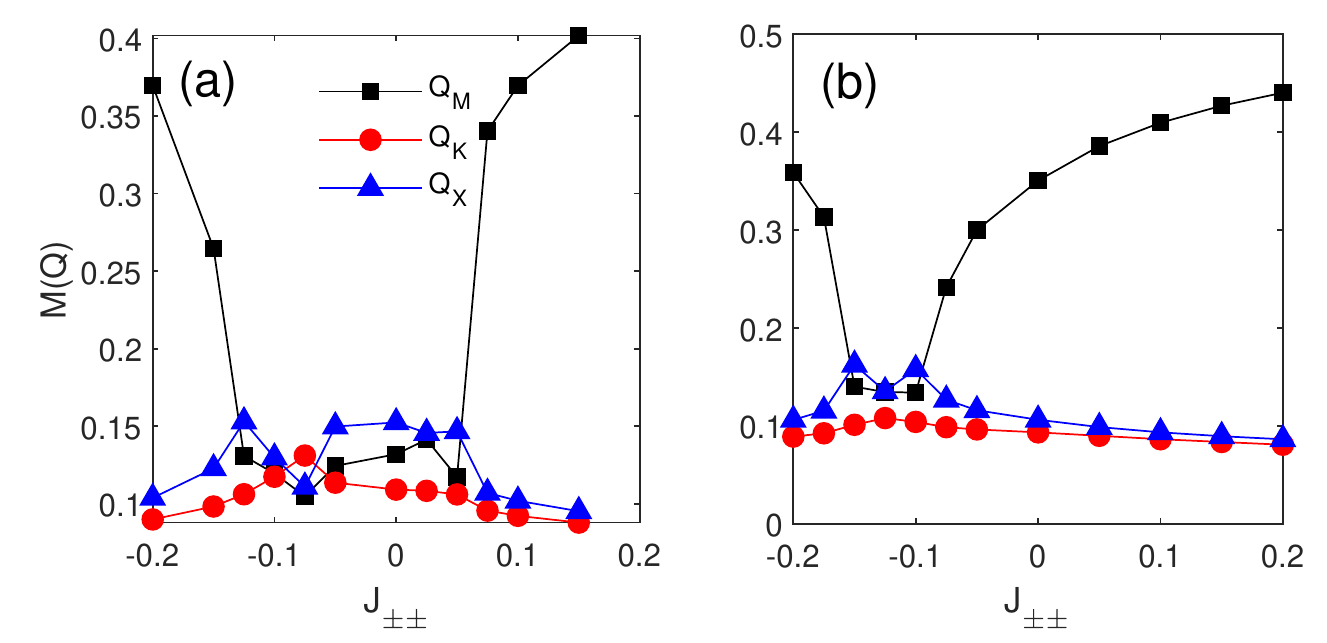}
	\caption{\label{SF line}
		 The order parameters $M(\mathbf{Q})$ as functions of $J_{\pm\pm}$ with (a) $J_{z\pm}=0.25$, and
		(b) $J_{z\pm}=0.35$. }
\end{figure}

To determine the phase boundary of the quantum model,
we scan the parameters in the $J_{\pm\pm}$-$J_{z\pm} $ plane with $\Delta\!=\!1$,
and calculate the order parameters.
Figure~\ref{SF line} depicts the order parameters
$M(\mathbf{Q})$  at ${\bf Q}$=$M$, $K$, $X$, as functions
of $J_{\pm\pm}$ for $J_{z\pm}=0.25$ [Fig.\ref{SF line}(a)] and
$J_{z\pm}=0.35$ [Fig.\ref{SF line}(b)], respectively.
The results are obtained on the 12$\times$12 lattice. These results suggest that the
intermediate region of the phase diagram is multi-\textbf{Q} phase, which can be further identified as a $Z_2$ vortex phase.
There are first-order like phase transitions from the $120^{\circ}$/ multi-\textbf{Q} phases to the stripe phases,
which are consistent to previous DMRG study \cite{zhu_topography_2018}.
The phase boundary between multi-\textbf{Q} phase and two stripe
phases is close to the boundary of classical results and classical spin-wave results \cite{maksimov_anisotropic-exchange_2019}.

\subsection{Additional results}
\label{sec:SFs}

\begin{figure}[tb!]
	\includegraphics[scale=0.38]{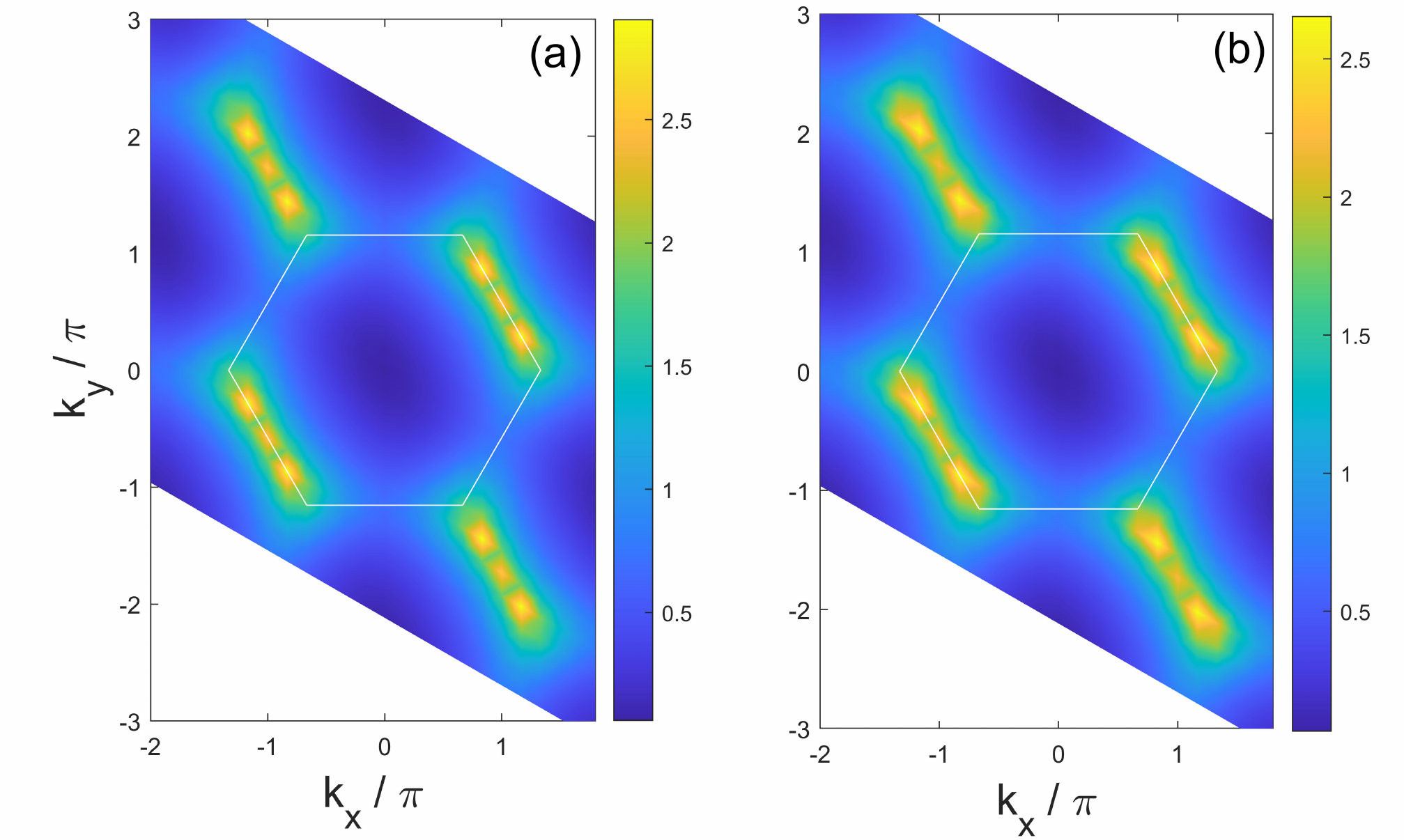}
	\caption{\label{fig:D8_D10}
	Comparing the total SSFs obtained by PEPS with (a) $D$=8 and (b) $D$=10 for $J_{\pm\pm}=-0.1,J_{z\pm}=0.3$.}
	
\end{figure}

Figure~\ref{fig:D8_D10} shows the total SSFs obtained by PEPS with $D$=8 and $D$=10 in the multi-\textbf{Q} phase, which
are essentially the same. Therefore, the PEPS with $D$=8 is enough to capture the essential physics of
the multi-\textbf{Q} state.

\begin{figure}[t!]
	\includegraphics[scale=0.40]{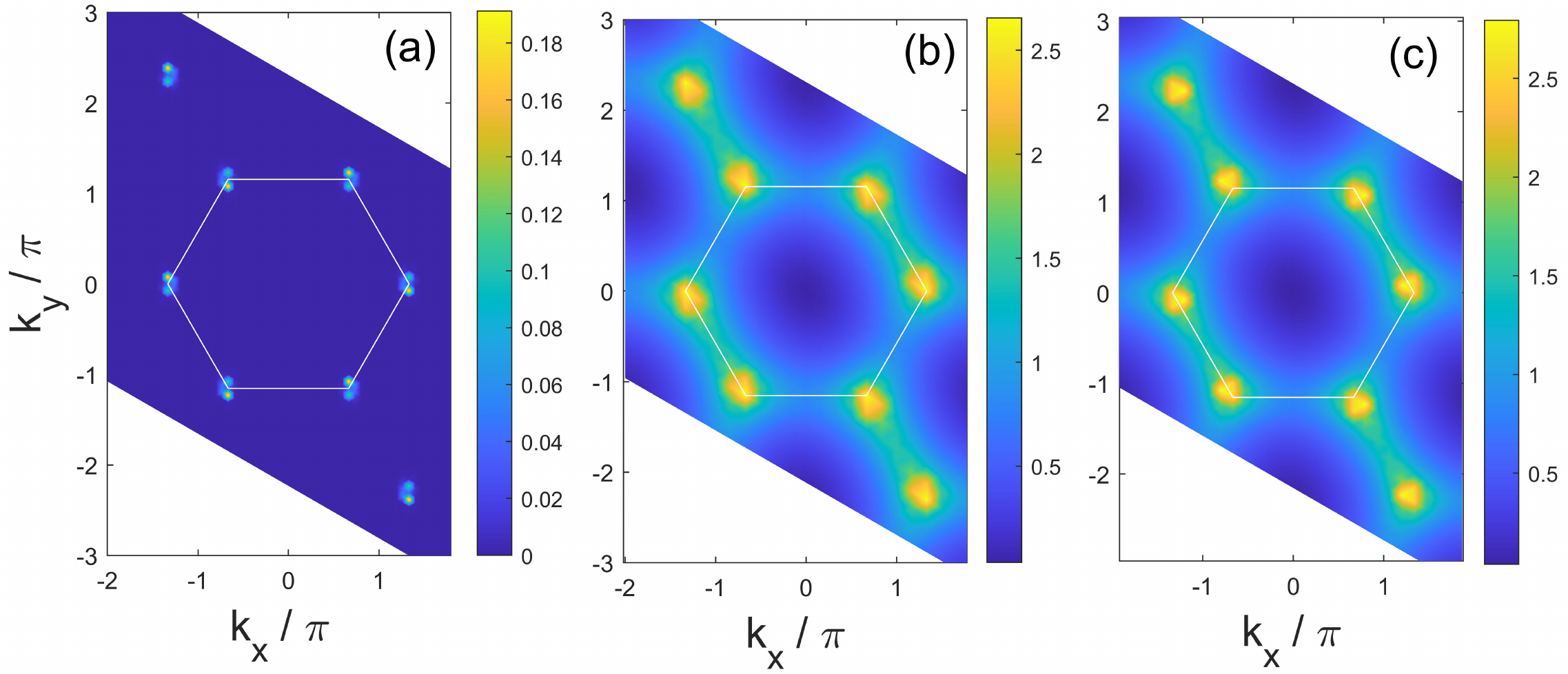}
		\caption{\label{fig:sf_m120} The total SSFs calculated at $J_{\pm\pm}=-0.025,J_{z\pm}=0.225$:
		(a) the SSF of the classical model, and (b),(c) the SSFs of the quantum model calculated at
the $12\times12$  and $15\times15$ lattices by PEPS, respectively.}
	
\end{figure}

Figure~\ref{fig:sf_m120} depicts the total SSFs calculated at $J_{\pm\pm}$=-0.025, and $J_{z\pm}$= 0.225.
Figure~\ref{fig:sf_m120}(a) shows the SSF of the classical model
calculated on a $30\times 30$ lattice, whose peaks move slightly away from the $K$ points.
The SSF suggests that the ground state is a modulated $120^{\circ}$ phase. Figure~\ref{fig:sf_m120}(b)(c) depict the SSFs of quantum model calculated by PEPS on the $12\times12$ and $15\times15$ lattices respectively. For the $12\times12$ lattice, the peaks of the SSF are located on the $K$ points, suggesting that the ground state is a $120^{\circ}$ phase.
However, the SSF of the $15\times15$ lattice is similar to that of the classical modulated $120^{\circ}$ phase, whose primary peaks move slightly away from the $K$ points, which means that the $120^{\circ}$ phase is unstable at this point. Therefore, the region of the $120^{\circ}$ phase should be smaller than that calculated on the $12\times12$ lattice in the thermodynamic limit.

\begin{thebibliography}{35}%
\makeatletter
\providecommand \@ifxundefined [1]{%
 \@ifx{#1\undefined}
}%
\providecommand \@ifnum [1]{%
 \ifnum #1\expandafter \@firstoftwo
 \else \expandafter \@secondoftwo
 \fi
}%
\providecommand \@ifx [1]{%
 \ifx #1\expandafter \@firstoftwo
 \else \expandafter \@secondoftwo
 \fi
}%
\providecommand \natexlab [1]{#1}%
\providecommand \enquote  [1]{``#1''}%
\providecommand \bibnamefont  [1]{#1}%
\providecommand \bibfnamefont [1]{#1}%
\providecommand \citenamefont [1]{#1}%
\providecommand \href@noop [0]{\@secondoftwo}%
\providecommand \href [0]{\begingroup \@sanitize@url \@href}%
\providecommand \@href[1]{\@@startlink{#1}\@@href}%
\providecommand \@@href[1]{\endgroup#1\@@endlink}%
\providecommand \@sanitize@url [0]{\catcode `\\12\catcode `\$12\catcode
  `\&12\catcode `\#12\catcode `\^12\catcode `\_12\catcode `\%12\relax}%
\providecommand \@@startlink[1]{}%
\providecommand \@@endlink[0]{}%
\providecommand \url  [0]{\begingroup\@sanitize@url \@url }%
\providecommand \@url [1]{\endgroup\@href {#1}{\urlprefix }}%
\providecommand \urlprefix  [0]{URL }%
\providecommand \Eprint [0]{\href }%
\providecommand \doibase [0]{https://doi.org/}%
\providecommand \selectlanguage [0]{\@gobble}%
\providecommand \bibinfo  [0]{\@secondoftwo}%
\providecommand \bibfield  [0]{\@secondoftwo}%
\providecommand \translation [1]{[#1]}%
\providecommand \BibitemOpen [0]{}%
\providecommand \bibitemStop [0]{}%
\providecommand \bibitemNoStop [0]{.\EOS\space}%
\providecommand \EOS [0]{\spacefactor3000\relax}%
\providecommand \BibitemShut  [1]{\csname bibitem#1\endcsname}%
\let\auto@bib@innerbib\@empty
\bibitem [{\citenamefont {Anderson}(1973)}]{anderson_resonating_1973}%
  \BibitemOpen
  \bibfield  {author} {\bibinfo {author} {\bibfnamefont {P.~W.}\ \bibnamefont
  {Anderson}},\ }\href
  {https://doi.org/https://doi.org/10.1016/0025-5408(73)90167-0} {\bibfield
  {journal} {\bibinfo  {journal} {Mater. Res. Bull.}\ }\textbf {\bibinfo
  {volume} {8}},\ \bibinfo {pages} {153 } (\bibinfo {year} {1973})}\BibitemShut
  {NoStop}%
\bibitem [{\citenamefont {Anderson}(1987)}]{anderson_resonating_1987}%
  \BibitemOpen
  \bibfield  {author} {\bibinfo {author} {\bibfnamefont {P.~W.}\ \bibnamefont
  {Anderson}},\ }\href {https://doi.org/10.1126/science.235.4793.1196}
  {\bibfield  {journal} {\bibinfo  {journal} {Science}\ }\textbf {\bibinfo
  {volume} {235}},\ \bibinfo {pages} {1196} (\bibinfo {year}
  {1987})}\BibitemShut {NoStop}%
\bibitem [{\citenamefont {Huse}\ and\ \citenamefont
  {Elser}(1988)}]{PhysRevLett.60.2531}%
  \BibitemOpen
  \bibfield  {author} {\bibinfo {author} {\bibfnamefont {D.~A.}\ \bibnamefont
  {Huse}}\ and\ \bibinfo {author} {\bibfnamefont {V.}~\bibnamefont {Elser}},\
  }\href {https://doi.org/10.1103/PhysRevLett.60.2531} {\bibfield  {journal}
  {\bibinfo  {journal} {Phys. Rev. Lett.}\ }\textbf {\bibinfo {volume} {60}},\
  \bibinfo {pages} {2531} (\bibinfo {year} {1988})}\BibitemShut {NoStop}%
\bibitem [{\citenamefont {Sindzingre}\ \emph {et~al.}(1994)\citenamefont
  {Sindzingre}, \citenamefont {Lecheminant},\ and\ \citenamefont
  {Lhuillier}}]{PhysRevB.50.3108}%
  \BibitemOpen
  \bibfield  {author} {\bibinfo {author} {\bibfnamefont {P.}~\bibnamefont
  {Sindzingre}}, \bibinfo {author} {\bibfnamefont {P.}~\bibnamefont
  {Lecheminant}},\ and\ \bibinfo {author} {\bibfnamefont {C.}~\bibnamefont
  {Lhuillier}},\ }\href {https://doi.org/10.1103/PhysRevB.50.3108} {\bibfield
  {journal} {\bibinfo  {journal} {Phys. Rev. B}\ }\textbf {\bibinfo {volume}
  {50}},\ \bibinfo {pages} {3108} (\bibinfo {year} {1994})}\BibitemShut
  {NoStop}%
\bibitem [{\citenamefont {Kitaev}(2006)}]{kitaev_anyons_2006}%
  \BibitemOpen
  \bibfield  {author} {\bibinfo {author} {\bibfnamefont {A.}~\bibnamefont
  {Kitaev}},\ }\href
  {https://doi.org/https://doi.org/10.1016/j.aop.2005.10.005} {\bibfield
  {journal} {\bibinfo  {journal} {Ann. Phys.}\ }\textbf {\bibinfo {volume}
  {321}},\ \bibinfo {pages} {2 } (\bibinfo {year} {2006})}\BibitemShut
  {NoStop}%
\bibitem [{\citenamefont {Li}\ \emph {et~al.}(2015{\natexlab{a}})\citenamefont
  {Li}, \citenamefont {Liao}, \citenamefont {Zhang}, \citenamefont {Li},
  \citenamefont {Jin}, \citenamefont {Ling}, \citenamefont {Zhang},
  \citenamefont {Zou}, \citenamefont {Pi}, \citenamefont {Yang}, \citenamefont
  {Wang}, \citenamefont {Wu},\ and\ \citenamefont {Zhang}}]{li_gapless_2015}%
  \BibitemOpen
  \bibfield  {author} {\bibinfo {author} {\bibfnamefont {Y.}~\bibnamefont
  {Li}}, \bibinfo {author} {\bibfnamefont {H.}~\bibnamefont {Liao}}, \bibinfo
  {author} {\bibfnamefont {Z.}~\bibnamefont {Zhang}}, \bibinfo {author}
  {\bibfnamefont {S.}~\bibnamefont {Li}}, \bibinfo {author} {\bibfnamefont
  {F.}~\bibnamefont {Jin}}, \bibinfo {author} {\bibfnamefont {L.}~\bibnamefont
  {Ling}}, \bibinfo {author} {\bibfnamefont {L.}~\bibnamefont {Zhang}},
  \bibinfo {author} {\bibfnamefont {Y.}~\bibnamefont {Zou}}, \bibinfo {author}
  {\bibfnamefont {L.}~\bibnamefont {Pi}}, \bibinfo {author} {\bibfnamefont
  {Z.}~\bibnamefont {Yang}}, \bibinfo {author} {\bibfnamefont {J.}~\bibnamefont
  {Wang}}, \bibinfo {author} {\bibfnamefont {Z.}~\bibnamefont {Wu}},\ and\
  \bibinfo {author} {\bibfnamefont {Q.}~\bibnamefont {Zhang}},\ }\href
  {https://doi.org/10.1038/srep16419} {\bibfield  {journal} {\bibinfo
  {journal} {Sci Rep}\ }\textbf {\bibinfo {volume} {5}},\ \bibinfo {pages}
  {16419} (\bibinfo {year} {2015}{\natexlab{a}})}\BibitemShut {NoStop}%
\bibitem [{\citenamefont {Li}\ \emph {et~al.}(2015{\natexlab{b}})\citenamefont
  {Li}, \citenamefont {Yu},\ and\ \citenamefont {Li}}]{Li_2015}%
  \BibitemOpen
  \bibfield  {author} {\bibinfo {author} {\bibfnamefont {K.}~\bibnamefont
  {Li}}, \bibinfo {author} {\bibfnamefont {S.-L.}\ \bibnamefont {Yu}},\ and\
  \bibinfo {author} {\bibfnamefont {J.-X.}\ \bibnamefont {Li}},\ }\href
  {https://doi.org/10.1088/1367-2630/17/4/043032} {\bibfield  {journal}
  {\bibinfo  {journal} {New J. Phys.}\ }\textbf {\bibinfo {volume} {17}},\
  \bibinfo {pages} {043032} (\bibinfo {year} {2015}{\natexlab{b}})}\BibitemShut
  {NoStop}%
\bibitem [{\citenamefont {Li}\ \emph {et~al.}(2015{\natexlab{c}})\citenamefont
  {Li}, \citenamefont {Chen}, \citenamefont {Tong}, \citenamefont {Pi},
  \citenamefont {Liu}, \citenamefont {Yang}, \citenamefont {Wang},\ and\
  \citenamefont {Zhang}}]{li_rare-earth_2015}%
  \BibitemOpen
  \bibfield  {author} {\bibinfo {author} {\bibfnamefont {Y.}~\bibnamefont
  {Li}}, \bibinfo {author} {\bibfnamefont {G.}~\bibnamefont {Chen}}, \bibinfo
  {author} {\bibfnamefont {W.}~\bibnamefont {Tong}}, \bibinfo {author}
  {\bibfnamefont {L.}~\bibnamefont {Pi}}, \bibinfo {author} {\bibfnamefont
  {J.}~\bibnamefont {Liu}}, \bibinfo {author} {\bibfnamefont {Z.}~\bibnamefont
  {Yang}}, \bibinfo {author} {\bibfnamefont {X.}~\bibnamefont {Wang}},\ and\
  \bibinfo {author} {\bibfnamefont {Q.}~\bibnamefont {Zhang}},\ }\href
  {https://doi.org/10.1103/PhysRevLett.115.167203} {\bibfield  {journal}
  {\bibinfo  {journal} {Phys. Rev. Lett.}\ }\textbf {\bibinfo {volume} {115}},\
  \bibinfo {pages} {167203} (\bibinfo {year} {2015}{\natexlab{c}})}\BibitemShut
  {NoStop}%
\bibitem [{\citenamefont {Li}\ \emph {et~al.}(2016{\natexlab{a}})\citenamefont
  {Li}, \citenamefont {Adroja}, \citenamefont {Biswas}, \citenamefont {Baker},
  \citenamefont {Zhang}, \citenamefont {Liu}, \citenamefont {Tsirlin},
  \citenamefont {Gegenwart},\ and\ \citenamefont {Zhang}}]{li_muon_2016}%
  \BibitemOpen
  \bibfield  {author} {\bibinfo {author} {\bibfnamefont {Y.}~\bibnamefont
  {Li}}, \bibinfo {author} {\bibfnamefont {D.}~\bibnamefont {Adroja}}, \bibinfo
  {author} {\bibfnamefont {P.~K.}\ \bibnamefont {Biswas}}, \bibinfo {author}
  {\bibfnamefont {P.~J.}\ \bibnamefont {Baker}}, \bibinfo {author}
  {\bibfnamefont {Q.}~\bibnamefont {Zhang}}, \bibinfo {author} {\bibfnamefont
  {J.}~\bibnamefont {Liu}}, \bibinfo {author} {\bibfnamefont {A.~A.}\
  \bibnamefont {Tsirlin}}, \bibinfo {author} {\bibfnamefont {P.}~\bibnamefont
  {Gegenwart}},\ and\ \bibinfo {author} {\bibfnamefont {Q.}~\bibnamefont
  {Zhang}},\ }\href {https://doi.org/10.1103/PhysRevLett.117.097201} {\bibfield
   {journal} {\bibinfo  {journal} {Phys. Rev. Lett.}\ }\textbf {\bibinfo
  {volume} {117}},\ \bibinfo {pages} {097201} (\bibinfo {year}
  {2016}{\natexlab{a}})}\BibitemShut {NoStop}%
\bibitem [{\citenamefont {Li}\ \emph {et~al.}(2017)\citenamefont {Li},
  \citenamefont {Adroja}, \citenamefont {Bewley}, \citenamefont {Voneshen},
  \citenamefont {Tsirlin}, \citenamefont {Gegenwart},\ and\ \citenamefont
  {Zhang}}]{li_crystalline_2017}%
  \BibitemOpen
  \bibfield  {author} {\bibinfo {author} {\bibfnamefont {Y.}~\bibnamefont
  {Li}}, \bibinfo {author} {\bibfnamefont {D.}~\bibnamefont {Adroja}}, \bibinfo
  {author} {\bibfnamefont {R.~I.}\ \bibnamefont {Bewley}}, \bibinfo {author}
  {\bibfnamefont {D.}~\bibnamefont {Voneshen}}, \bibinfo {author}
  {\bibfnamefont {A.~A.}\ \bibnamefont {Tsirlin}}, \bibinfo {author}
  {\bibfnamefont {P.}~\bibnamefont {Gegenwart}},\ and\ \bibinfo {author}
  {\bibfnamefont {Q.}~\bibnamefont {Zhang}},\ }\href
  {https://doi.org/10.1103/PhysRevLett.118.107202} {\bibfield  {journal}
  {\bibinfo  {journal} {Phys. Rev. Lett.}\ }\textbf {\bibinfo {volume} {118}},\
  \bibinfo {pages} {107202} (\bibinfo {year} {2017})}\BibitemShut {NoStop}%
\bibitem [{\citenamefont {Shen}\ \emph {et~al.}(2016)\citenamefont {Shen},
  \citenamefont {Li}, \citenamefont {Wo}, \citenamefont {Li}, \citenamefont
  {Shen}, \citenamefont {Pan}, \citenamefont {Wang}, \citenamefont {Walker},
  \citenamefont {Steffens}, \citenamefont {Boehm}, \citenamefont {Hao},
  \citenamefont {Quintero-Castro}, \citenamefont {Harriger}, \citenamefont
  {Frontzek}, \citenamefont {Hao}, \citenamefont {Meng}, \citenamefont {Zhang},
  \citenamefont {Chen},\ and\ \citenamefont {Zhao}}]{shen_evidence_2016}%
  \BibitemOpen
  \bibfield  {author} {\bibinfo {author} {\bibfnamefont {Y.}~\bibnamefont
  {Shen}}, \bibinfo {author} {\bibfnamefont {Y.-D.}\ \bibnamefont {Li}},
  \bibinfo {author} {\bibfnamefont {H.}~\bibnamefont {Wo}}, \bibinfo {author}
  {\bibfnamefont {Y.}~\bibnamefont {Li}}, \bibinfo {author} {\bibfnamefont
  {S.}~\bibnamefont {Shen}}, \bibinfo {author} {\bibfnamefont {B.}~\bibnamefont
  {Pan}}, \bibinfo {author} {\bibfnamefont {Q.}~\bibnamefont {Wang}}, \bibinfo
  {author} {\bibfnamefont {H.~C.}\ \bibnamefont {Walker}}, \bibinfo {author}
  {\bibfnamefont {P.}~\bibnamefont {Steffens}}, \bibinfo {author}
  {\bibfnamefont {M.}~\bibnamefont {Boehm}}, \bibinfo {author} {\bibfnamefont
  {Y.}~\bibnamefont {Hao}}, \bibinfo {author} {\bibfnamefont {D.~L.}\
  \bibnamefont {Quintero-Castro}}, \bibinfo {author} {\bibfnamefont {L.~W.}\
  \bibnamefont {Harriger}}, \bibinfo {author} {\bibfnamefont {M.~D.}\
  \bibnamefont {Frontzek}}, \bibinfo {author} {\bibfnamefont {L.}~\bibnamefont
  {Hao}}, \bibinfo {author} {\bibfnamefont {S.}~\bibnamefont {Meng}}, \bibinfo
  {author} {\bibfnamefont {Q.}~\bibnamefont {Zhang}}, \bibinfo {author}
  {\bibfnamefont {G.}~\bibnamefont {Chen}},\ and\ \bibinfo {author}
  {\bibfnamefont {J.}~\bibnamefont {Zhao}},\ }\href
  {http://dx.doi.org/10.1038/nature20614} {\bibfield  {journal} {\bibinfo
  {journal} {Nature}\ }\textbf {\bibinfo {volume} {540}},\ \bibinfo {pages}
  {559} (\bibinfo {year} {2016})}\BibitemShut {NoStop}%
\bibitem [{\citenamefont {Paddison}\ \emph {et~al.}(2016)\citenamefont
  {Paddison}, \citenamefont {Daum}, \citenamefont {Dun}, \citenamefont
  {Ehlers}, \citenamefont {Liu}, \citenamefont {Stone}, \citenamefont {Zhou},\
  and\ \citenamefont {Mourigal}}]{paddison_continuous_2016}%
  \BibitemOpen
  \bibfield  {author} {\bibinfo {author} {\bibfnamefont {J.~A.~M.}\
  \bibnamefont {Paddison}}, \bibinfo {author} {\bibfnamefont {M.}~\bibnamefont
  {Daum}}, \bibinfo {author} {\bibfnamefont {Z.}~\bibnamefont {Dun}}, \bibinfo
  {author} {\bibfnamefont {G.}~\bibnamefont {Ehlers}}, \bibinfo {author}
  {\bibfnamefont {Y.}~\bibnamefont {Liu}}, \bibinfo {author} {\bibfnamefont
  {M.~B.}\ \bibnamefont {Stone}}, \bibinfo {author} {\bibfnamefont
  {H.}~\bibnamefont {Zhou}},\ and\ \bibinfo {author} {\bibfnamefont
  {M.}~\bibnamefont {Mourigal}},\ }\href {https://doi.org/10.1038/nphys3971}
  {\bibfield  {journal} {\bibinfo  {journal} {Nat. Phys.}\ }\textbf {\bibinfo
  {volume} {13}},\ \bibinfo {pages} {117} (\bibinfo {year} {2016})}\BibitemShut
  {NoStop}%
\bibitem [{\citenamefont {Li}\ \emph {et~al.}(2019{\natexlab{a}})\citenamefont
  {Li}, \citenamefont {Bachus}, \citenamefont {Liu}, \citenamefont
  {Radelytskyi}, \citenamefont {Bertin}, \citenamefont {Schneidewind},
  \citenamefont {Tokiwa}, \citenamefont {Tsirlin},\ and\ \citenamefont
  {Gegenwart}}]{li_rearrangement_2019}%
  \BibitemOpen
  \bibfield  {author} {\bibinfo {author} {\bibfnamefont {Y.}~\bibnamefont
  {Li}}, \bibinfo {author} {\bibfnamefont {S.}~\bibnamefont {Bachus}}, \bibinfo
  {author} {\bibfnamefont {B.}~\bibnamefont {Liu}}, \bibinfo {author}
  {\bibfnamefont {I.}~\bibnamefont {Radelytskyi}}, \bibinfo {author}
  {\bibfnamefont {A.}~\bibnamefont {Bertin}}, \bibinfo {author} {\bibfnamefont
  {A.}~\bibnamefont {Schneidewind}}, \bibinfo {author} {\bibfnamefont
  {Y.}~\bibnamefont {Tokiwa}}, \bibinfo {author} {\bibfnamefont {A.~A.}\
  \bibnamefont {Tsirlin}},\ and\ \bibinfo {author} {\bibfnamefont
  {P.}~\bibnamefont {Gegenwart}},\ }\href
  {https://doi.org/10.1103/PhysRevLett.122.137201} {\bibfield  {journal}
  {\bibinfo  {journal} {Phys. Rev. Lett.}\ }\textbf {\bibinfo {volume} {122}},\
  \bibinfo {pages} {137201} (\bibinfo {year} {2019}{\natexlab{a}})}\BibitemShut
  {NoStop}%
\bibitem [{\citenamefont {Liu}\ \emph {et~al.}(2018)\citenamefont {Liu},
  \citenamefont {Zhang}, \citenamefont {Ji}, \citenamefont {Liu}, \citenamefont
  {Li}, \citenamefont {Wang}, \citenamefont {Lei}, \citenamefont {Chen},\ and\
  \citenamefont {Zhang}}]{liu_rare-earth_2018}%
  \BibitemOpen
  \bibfield  {author} {\bibinfo {author} {\bibfnamefont {W.}~\bibnamefont
  {Liu}}, \bibinfo {author} {\bibfnamefont {Z.}~\bibnamefont {Zhang}}, \bibinfo
  {author} {\bibfnamefont {J.}~\bibnamefont {Ji}}, \bibinfo {author}
  {\bibfnamefont {Y.}~\bibnamefont {Liu}}, \bibinfo {author} {\bibfnamefont
  {J.}~\bibnamefont {Li}}, \bibinfo {author} {\bibfnamefont {X.}~\bibnamefont
  {Wang}}, \bibinfo {author} {\bibfnamefont {H.}~\bibnamefont {Lei}}, \bibinfo
  {author} {\bibfnamefont {G.}~\bibnamefont {Chen}},\ and\ \bibinfo {author}
  {\bibfnamefont {Q.}~\bibnamefont {Zhang}},\ }\href
  {https://doi.org/10.1088/0256-307X/35/11/117501} {\bibfield  {journal}
  {\bibinfo  {journal} {Chinese Phys. Lett.}\ }\textbf {\bibinfo {volume}
  {35}},\ \bibinfo {pages} {117501} (\bibinfo {year} {2018})}\BibitemShut
  {NoStop}%
\bibitem [{\citenamefont {Bordelon}\ \emph {et~al.}(2019)\citenamefont
  {Bordelon}, \citenamefont {Kenney}, \citenamefont {Liu}, \citenamefont
  {Hogan}, \citenamefont {Posthuma}, \citenamefont {Kavand}, \citenamefont
  {Lyu}, \citenamefont {Sherwin}, \citenamefont {Butch}, \citenamefont {Brown},
  \citenamefont {Graf}, \citenamefont {Balents},\ and\ \citenamefont
  {Wilson}}]{bordelon_field-tunable_2019}%
  \BibitemOpen
  \bibfield  {author} {\bibinfo {author} {\bibfnamefont {M.~M.}\ \bibnamefont
  {Bordelon}}, \bibinfo {author} {\bibfnamefont {E.}~\bibnamefont {Kenney}},
  \bibinfo {author} {\bibfnamefont {C.}~\bibnamefont {Liu}}, \bibinfo {author}
  {\bibfnamefont {T.}~\bibnamefont {Hogan}}, \bibinfo {author} {\bibfnamefont
  {L.}~\bibnamefont {Posthuma}}, \bibinfo {author} {\bibfnamefont
  {M.}~\bibnamefont {Kavand}}, \bibinfo {author} {\bibfnamefont
  {Y.}~\bibnamefont {Lyu}}, \bibinfo {author} {\bibfnamefont {M.}~\bibnamefont
  {Sherwin}}, \bibinfo {author} {\bibfnamefont {N.~P.}\ \bibnamefont {Butch}},
  \bibinfo {author} {\bibfnamefont {C.}~\bibnamefont {Brown}}, \bibinfo
  {author} {\bibfnamefont {M.~J.}\ \bibnamefont {Graf}}, \bibinfo {author}
  {\bibfnamefont {L.}~\bibnamefont {Balents}},\ and\ \bibinfo {author}
  {\bibfnamefont {S.~D.}\ \bibnamefont {Wilson}},\ }\href
  {https://doi.org/10.1038/s41567-019-0594-5} {\bibfield  {journal} {\bibinfo
  {journal} {Nat. Phys.}\ }\textbf {\bibinfo {volume} {15}},\ \bibinfo {pages}
  {1058} (\bibinfo {year} {2019})}\BibitemShut {NoStop}%
\bibitem [{\citenamefont {Li}\ \emph {et~al.}(2016{\natexlab{b}})\citenamefont
  {Li}, \citenamefont {Wang},\ and\ \citenamefont
  {Chen}}]{li_anisotropic_2016}%
  \BibitemOpen
  \bibfield  {author} {\bibinfo {author} {\bibfnamefont {Y.-D.}\ \bibnamefont
  {Li}}, \bibinfo {author} {\bibfnamefont {X.}~\bibnamefont {Wang}},\ and\
  \bibinfo {author} {\bibfnamefont {G.}~\bibnamefont {Chen}},\ }\href
  {https://doi.org/10.1103/PhysRevB.94.035107} {\bibfield  {journal} {\bibinfo
  {journal} {Phys. Rev. B}\ }\textbf {\bibinfo {volume} {94}},\ \bibinfo
  {pages} {035107} (\bibinfo {year} {2016}{\natexlab{b}})}\BibitemShut
  {NoStop}%
\bibitem [{\citenamefont {Maksimov}\ \emph {et~al.}(2019)\citenamefont
  {Maksimov}, \citenamefont {Zhu}, \citenamefont {White},\ and\ \citenamefont
  {Chernyshev}}]{maksimov_anisotropic-exchange_2019}%
  \BibitemOpen
  \bibfield  {author} {\bibinfo {author} {\bibfnamefont {P.~A.}\ \bibnamefont
  {Maksimov}}, \bibinfo {author} {\bibfnamefont {Z.}~\bibnamefont {Zhu}},
  \bibinfo {author} {\bibfnamefont {S.~R.}\ \bibnamefont {White}},\ and\
  \bibinfo {author} {\bibfnamefont {A.~L.}\ \bibnamefont {Chernyshev}},\ }\href
  {https://doi.org/10.1103/PhysRevX.9.021017} {\bibfield  {journal} {\bibinfo
  {journal} {Phys. Rev. X}\ }\textbf {\bibinfo {volume} {9}},\ \bibinfo {pages}
  {021017} (\bibinfo {year} {2019})}\BibitemShut {NoStop}%
\bibitem [{\citenamefont {Becker}\ \emph {et~al.}(2015)\citenamefont {Becker},
  \citenamefont {Hermanns}, \citenamefont {Bauer}, \citenamefont {Garst},\ and\
  \citenamefont {Trebst}}]{becker_spin-orbit_2015}%
  \BibitemOpen
  \bibfield  {author} {\bibinfo {author} {\bibfnamefont {M.}~\bibnamefont
  {Becker}}, \bibinfo {author} {\bibfnamefont {M.}~\bibnamefont {Hermanns}},
  \bibinfo {author} {\bibfnamefont {B.}~\bibnamefont {Bauer}}, \bibinfo
  {author} {\bibfnamefont {M.}~\bibnamefont {Garst}},\ and\ \bibinfo {author}
  {\bibfnamefont {S.}~\bibnamefont {Trebst}},\ }\href
  {https://doi.org/10.1103/PhysRevB.91.155135} {\bibfield  {journal} {\bibinfo
  {journal} {Phys. Rev. B}\ }\textbf {\bibinfo {volume} {91}},\ \bibinfo
  {pages} {155135} (\bibinfo {year} {2015})}\BibitemShut {NoStop}%
\bibitem [{\citenamefont {Rousochatzakis}\ \emph {et~al.}(2016)\citenamefont
  {Rousochatzakis}, \citenamefont {R\"{o}ssler}, \citenamefont {van~den
  Brink},\ and\ \citenamefont {Daghofer}}]{rousochatzakis_kitaev_2016}%
  \BibitemOpen
  \bibfield  {author} {\bibinfo {author} {\bibfnamefont {I.}~\bibnamefont
  {Rousochatzakis}}, \bibinfo {author} {\bibfnamefont {U.~K.}\ \bibnamefont
  {R\"{o}ssler}}, \bibinfo {author} {\bibfnamefont {J.}~\bibnamefont {van~den
  Brink}},\ and\ \bibinfo {author} {\bibfnamefont {M.}~\bibnamefont
  {Daghofer}},\ }\href {https://doi.org/10.1103/PhysRevB.93.104417} {\bibfield
  {journal} {\bibinfo  {journal} {Phys. Rev. B}\ }\textbf {\bibinfo {volume}
  {93}},\ \bibinfo {pages} {104417} (\bibinfo {year} {2016})}\BibitemShut
  {NoStop}%
\bibitem [{\citenamefont {Liu}\ \emph {et~al.}(2016)\citenamefont {Liu},
  \citenamefont {Yu},\ and\ \citenamefont {Wang}}]{liu_semiclassical_2016}%
  \BibitemOpen
  \bibfield  {author} {\bibinfo {author} {\bibfnamefont {C.}~\bibnamefont
  {Liu}}, \bibinfo {author} {\bibfnamefont {R.}~\bibnamefont {Yu}},\ and\
  \bibinfo {author} {\bibfnamefont {X.}~\bibnamefont {Wang}},\ }\href
  {https://doi.org/10.1103/PhysRevB.94.174424} {\bibfield  {journal} {\bibinfo
  {journal} {Phys. Rev. B}\ }\textbf {\bibinfo {volume} {94}},\ \bibinfo
  {pages} {174424} (\bibinfo {year} {2016})}\BibitemShut {NoStop}%
\bibitem [{\citenamefont {Kos}\ and\ \citenamefont
  {Punk}(2017)}]{PhysRevB.95.024421}%
  \BibitemOpen
  \bibfield  {author} {\bibinfo {author} {\bibfnamefont {P.}~\bibnamefont
  {Kos}}\ and\ \bibinfo {author} {\bibfnamefont {M.}~\bibnamefont {Punk}},\
  }\href {https://doi.org/10.1103/PhysRevB.95.024421} {\bibfield  {journal}
  {\bibinfo  {journal} {Phys. Rev. B}\ }\textbf {\bibinfo {volume} {95}},\
  \bibinfo {pages} {024421} (\bibinfo {year} {2017})}\BibitemShut {NoStop}%
\bibitem [{\citenamefont {Wu}\ \emph {et~al.}(2020)\citenamefont {Wu},
  \citenamefont {Yao},\ and\ \citenamefont {Wu}}]{wu_exact_2020}%
  \BibitemOpen
  \bibfield  {author} {\bibinfo {author} {\bibfnamefont {M.}~\bibnamefont
  {Wu}}, \bibinfo {author} {\bibfnamefont {D.-X.}\ \bibnamefont {Yao}},\ and\
  \bibinfo {author} {\bibfnamefont {H.-Q.}\ \bibnamefont {Wu}},\ }\href
  {http://arxiv.org/abs/2008.08751} {\bibfield  {journal} {\bibinfo  {journal}
  {arXiv:2008.08751 [cond-mat]}\ } (\bibinfo {year} {2020})}\BibitemShut
  {NoStop}%
\bibitem [{\citenamefont {Zhu}\ \emph {et~al.}(2018)\citenamefont {Zhu},
  \citenamefont {Maksimov}, \citenamefont {White},\ and\ \citenamefont
  {Chernyshev}}]{zhu_topography_2018}%
  \BibitemOpen
  \bibfield  {author} {\bibinfo {author} {\bibfnamefont {Z.}~\bibnamefont
  {Zhu}}, \bibinfo {author} {\bibfnamefont {P.~A.}\ \bibnamefont {Maksimov}},
  \bibinfo {author} {\bibfnamefont {S.~R.}\ \bibnamefont {White}},\ and\
  \bibinfo {author} {\bibfnamefont {A.~L.}\ \bibnamefont {Chernyshev}},\ }\href
  {https://doi.org/10.1103/PhysRevLett.120.207203} {\bibfield  {journal}
  {\bibinfo  {journal} {Phys. Rev. Lett.}\ }\textbf {\bibinfo {volume} {120}},\
  \bibinfo {pages} {207203} (\bibinfo {year} {2018})}\BibitemShut {NoStop}%
\bibitem [{\citenamefont {Luo}\ \emph {et~al.}(2017)\citenamefont {Luo},
  \citenamefont {Hu}, \citenamefont {Xi}, \citenamefont {Zhao},\ and\
  \citenamefont {Wang}}]{luo_ground-state_2017}%
  \BibitemOpen
  \bibfield  {author} {\bibinfo {author} {\bibfnamefont {Q.}~\bibnamefont
  {Luo}}, \bibinfo {author} {\bibfnamefont {S.}~\bibnamefont {Hu}}, \bibinfo
  {author} {\bibfnamefont {B.}~\bibnamefont {Xi}}, \bibinfo {author}
  {\bibfnamefont {J.}~\bibnamefont {Zhao}},\ and\ \bibinfo {author}
  {\bibfnamefont {X.}~\bibnamefont {Wang}},\ }\href
  {https://doi.org/10.1103/PhysRevB.95.165110} {\bibfield  {journal} {\bibinfo
  {journal} {Phys. Rev. B}\ }\textbf {\bibinfo {volume} {95}},\ \bibinfo
  {pages} {165110} (\bibinfo {year} {2017})}\BibitemShut {NoStop}%
\bibitem [{\citenamefont {Shinjo}\ \emph {et~al.}(2016)\citenamefont {Shinjo},
  \citenamefont {Sota}, \citenamefont {Yunoki}, \citenamefont {Totsuka},\ and\
  \citenamefont {Tohyama}}]{shinjo_density-matrix_2016}%
  \BibitemOpen
  \bibfield  {author} {\bibinfo {author} {\bibfnamefont {K.}~\bibnamefont
  {Shinjo}}, \bibinfo {author} {\bibfnamefont {S.}~\bibnamefont {Sota}},
  \bibinfo {author} {\bibfnamefont {S.}~\bibnamefont {Yunoki}}, \bibinfo
  {author} {\bibfnamefont {K.}~\bibnamefont {Totsuka}},\ and\ \bibinfo {author}
  {\bibfnamefont {T.}~\bibnamefont {Tohyama}},\ }\href
  {https://doi.org/10.7566/JPSJ.85.114710} {\bibfield  {journal} {\bibinfo
  {journal} {J. Phys. Soc. Jpn.}\ }\textbf {\bibinfo {volume} {85}},\ \bibinfo
  {pages} {114710} (\bibinfo {year} {2016})}\BibitemShut {NoStop}%
\bibitem [{\citenamefont {Zheng}\ \emph {et~al.}(2021)\citenamefont {Zheng},
  \citenamefont {Shen}, \citenamefont {Zhang},\ and\ \citenamefont
  {He}}]{zheng2021firstprinciples}%
  \BibitemOpen
  \bibfield  {author} {\bibinfo {author} {\bibfnamefont {D.-Y.}\ \bibnamefont
  {Zheng}}, \bibinfo {author} {\bibfnamefont {Z.-X.}\ \bibnamefont {Shen}},
  \bibinfo {author} {\bibfnamefont {M.}~\bibnamefont {Zhang}},\ and\ \bibinfo
  {author} {\bibfnamefont {L.}~\bibnamefont {He}},\ }\href
  {https://arxiv.org/abs/2104.09739} {\bibfield  {journal} {\bibinfo  {journal}
  {arXiv:2104.09739 [cond-mat.mtrl-sci]}\ } (\bibinfo {year}
  {2021})}\BibitemShut {NoStop}%
\bibitem [{\citenamefont {Swendsen}\ and\ \citenamefont
  {Wang}(1986)}]{PhysRevLett.57.2607}%
  \BibitemOpen
  \bibfield  {author} {\bibinfo {author} {\bibfnamefont {R.~H.}\ \bibnamefont
  {Swendsen}}\ and\ \bibinfo {author} {\bibfnamefont {J.-S.}\ \bibnamefont
  {Wang}},\ }\href {https://doi.org/10.1103/PhysRevLett.57.2607} {\bibfield
  {journal} {\bibinfo  {journal} {Phys. Rev. Lett.}\ }\textbf {\bibinfo
  {volume} {57}},\ \bibinfo {pages} {2607} (\bibinfo {year}
  {1986})}\BibitemShut {NoStop}%
\bibitem [{\citenamefont {Liu}\ \emph {et~al.}(2015)\citenamefont {Liu},
  \citenamefont {Wang}, \citenamefont {Li}, \citenamefont {Lao}, \citenamefont
  {Han}, \citenamefont {Guo}, \citenamefont {Zhao},\ and\ \citenamefont
  {He}}]{liu_replica_2015}%
  \BibitemOpen
  \bibfield  {author} {\bibinfo {author} {\bibfnamefont {W.}~\bibnamefont
  {Liu}}, \bibinfo {author} {\bibfnamefont {C.}~\bibnamefont {Wang}}, \bibinfo
  {author} {\bibfnamefont {Y.}~\bibnamefont {Li}}, \bibinfo {author}
  {\bibfnamefont {Y.}~\bibnamefont {Lao}}, \bibinfo {author} {\bibfnamefont
  {Y.}~\bibnamefont {Han}}, \bibinfo {author} {\bibfnamefont {G.-C.}\
  \bibnamefont {Guo}}, \bibinfo {author} {\bibfnamefont {Y.-H.}\ \bibnamefont
  {Zhao}},\ and\ \bibinfo {author} {\bibfnamefont {L.}~\bibnamefont {He}},\
  }\href {https://doi.org/10.1088/0953-8984/27/8/085601} {\bibfield  {journal}
  {\bibinfo  {journal} {J. Phys.: Condens. Matter}\ }\textbf {\bibinfo {volume}
  {27}},\ \bibinfo {pages} {085601} (\bibinfo {year} {2015})}\BibitemShut
  {NoStop}%
\bibitem [{\citenamefont {Wang}\ \emph {et~al.}(2017)\citenamefont {Wang},
  \citenamefont {Gong}, \citenamefont {Han}, \citenamefont {Guo},\ and\
  \citenamefont {He}}]{PhysRevB.96.115119}%
  \BibitemOpen
  \bibfield  {author} {\bibinfo {author} {\bibfnamefont {C.}~\bibnamefont
  {Wang}}, \bibinfo {author} {\bibfnamefont {M.}~\bibnamefont {Gong}}, \bibinfo
  {author} {\bibfnamefont {Y.}~\bibnamefont {Han}}, \bibinfo {author}
  {\bibfnamefont {G.}~\bibnamefont {Guo}},\ and\ \bibinfo {author}
  {\bibfnamefont {L.}~\bibnamefont {He}},\ }\href
  {https://doi.org/10.1103/PhysRevB.96.115119} {\bibfield  {journal} {\bibinfo
  {journal} {Phys. Rev. B}\ }\textbf {\bibinfo {volume} {96}},\ \bibinfo
  {pages} {115119} (\bibinfo {year} {2017})}\BibitemShut {NoStop}%
\bibitem [{\citenamefont {Jiang}\ \emph {et~al.}(2008)\citenamefont {Jiang},
  \citenamefont {Weng},\ and\ \citenamefont {Xiang}}]{PhysRevLett.101.090603}%
  \BibitemOpen
  \bibfield  {author} {\bibinfo {author} {\bibfnamefont {H.~C.}\ \bibnamefont
  {Jiang}}, \bibinfo {author} {\bibfnamefont {Z.~Y.}\ \bibnamefont {Weng}},\
  and\ \bibinfo {author} {\bibfnamefont {T.}~\bibnamefont {Xiang}},\ }\href
  {https://doi.org/10.1103/PhysRevLett.101.090603} {\bibfield  {journal}
  {\bibinfo  {journal} {Phys. Rev. Lett.}\ }\textbf {\bibinfo {volume} {101}},\
  \bibinfo {pages} {090603} (\bibinfo {year} {2008})}\BibitemShut {NoStop}%
\bibitem [{\citenamefont {Liu}\ \emph {et~al.}(2017)\citenamefont {Liu},
  \citenamefont {Dong}, \citenamefont {Han}, \citenamefont {Guo},\ and\
  \citenamefont {He}}]{PhysRevB.95.195154}%
  \BibitemOpen
  \bibfield  {author} {\bibinfo {author} {\bibfnamefont {W.-Y.}\ \bibnamefont
  {Liu}}, \bibinfo {author} {\bibfnamefont {S.-J.}\ \bibnamefont {Dong}},
  \bibinfo {author} {\bibfnamefont {Y.-J.}\ \bibnamefont {Han}}, \bibinfo
  {author} {\bibfnamefont {G.-C.}\ \bibnamefont {Guo}},\ and\ \bibinfo {author}
  {\bibfnamefont {L.}~\bibnamefont {He}},\ }\href
  {https://doi.org/10.1103/PhysRevB.95.195154} {\bibfield  {journal} {\bibinfo
  {journal} {Phys. Rev. B}\ }\textbf {\bibinfo {volume} {95}},\ \bibinfo
  {pages} {195154} (\bibinfo {year} {2017})}\BibitemShut {NoStop}%
\bibitem [{\citenamefont {Li}\ \emph {et~al.}(2019{\natexlab{b}})\citenamefont
  {Li}, \citenamefont {Perkins},\ and\ \citenamefont
  {Rousochatzakis}}]{li_collective_2019}%
  \BibitemOpen
  \bibfield  {author} {\bibinfo {author} {\bibfnamefont {M.}~\bibnamefont
  {Li}}, \bibinfo {author} {\bibfnamefont {N.~B.}\ \bibnamefont {Perkins}},\
  and\ \bibinfo {author} {\bibfnamefont {I.}~\bibnamefont {Rousochatzakis}},\
  }\href {https://doi.org/10.1103/PhysRevResearch.1.013002} {\bibfield
  {journal} {\bibinfo  {journal} {Phys. Rev. Research}\ }\textbf {\bibinfo
  {volume} {1}},\ \bibinfo {pages} {013002} (\bibinfo {year}
  {2019}{\natexlab{b}})}\BibitemShut {NoStop}%
\bibitem [{\citenamefont {Seabrook}\ \emph {et~al.}(2020)\citenamefont
  {Seabrook}, \citenamefont {Baez},\ and\ \citenamefont
  {Reuther}}]{PhysRevB.101.174443}%
  \BibitemOpen
  \bibfield  {author} {\bibinfo {author} {\bibfnamefont {E.}~\bibnamefont
  {Seabrook}}, \bibinfo {author} {\bibfnamefont {M.~L.}\ \bibnamefont {Baez}},\
  and\ \bibinfo {author} {\bibfnamefont {J.}~\bibnamefont {Reuther}},\ }\href
  {https://doi.org/10.1103/PhysRevB.101.174443} {\bibfield  {journal} {\bibinfo
   {journal} {Phys. Rev. B}\ }\textbf {\bibinfo {volume} {101}},\ \bibinfo
  {pages} {174443} (\bibinfo {year} {2020})}\BibitemShut {NoStop}%
\bibitem [{\citenamefont {Wang}\ \emph {et~al.}(2021)\citenamefont {Wang},
  \citenamefont {Qi}, \citenamefont {Xi}, \citenamefont {Wang}, \citenamefont
  {Yu},\ and\ \citenamefont {Li}}]{wang_comprehensive_2020}%
  \BibitemOpen
  \bibfield  {author} {\bibinfo {author} {\bibfnamefont {S.}~\bibnamefont
  {Wang}}, \bibinfo {author} {\bibfnamefont {Z.}~\bibnamefont {Qi}}, \bibinfo
  {author} {\bibfnamefont {B.}~\bibnamefont {Xi}}, \bibinfo {author}
  {\bibfnamefont {W.}~\bibnamefont {Wang}}, \bibinfo {author} {\bibfnamefont
  {S.-L.}\ \bibnamefont {Yu}},\ and\ \bibinfo {author} {\bibfnamefont {J.-X.}\
  \bibnamefont {Li}},\ }\href {https://doi.org/10.1103/PhysRevB.103.054410}
  {\bibfield  {journal} {\bibinfo  {journal} {Phys. Rev. B}\ }\textbf {\bibinfo
  {volume} {103}},\ \bibinfo {pages} {054410} (\bibinfo {year}
  {2021})}\BibitemShut {NoStop}%
\bibitem [{\citenamefont {Rau}\ and\ \citenamefont
  {Gingras}(2018)}]{PhysRevB.98.054408}%
  \BibitemOpen
  \bibfield  {author} {\bibinfo {author} {\bibfnamefont {J.~G.}\ \bibnamefont
  {Rau}}\ and\ \bibinfo {author} {\bibfnamefont {M.~J.~P.}\ \bibnamefont
  {Gingras}},\ }\href {https://doi.org/10.1103/PhysRevB.98.054408} {\bibfield
  {journal} {\bibinfo  {journal} {Phys. Rev. B}\ }\textbf {\bibinfo {volume}
  {98}},\ \bibinfo {pages} {054408} (\bibinfo {year} {2018})}\BibitemShut
  {NoStop}%
\end{thebibliography}
%

\end{document}